\documentclass[openacc]{rstransa}
\usepackage{amsfonts}
\usepackage{amscd}
\usepackage{bm}
\usepackage{amsmath}
\usepackage{amssymb}
\usepackage{esint}
\newcommand{\be}{\begin{equation}}
\newcommand{\ee}{\end{equation}}
\newcommand{\ber}{\begin{eqnarray}}
\newcommand{\eer}{\end{eqnarray}}
\newcommand{\sgn}{\mbox{\it sgn}}
\def\Tr#1{\mbox{Tr}\Big\{#1\Big\}}
\renewcommand{\Im}{\mbox{Im\,}}
\renewcommand{\Re}{\mbox{Re\,}}
\newcommand{\whtaux}{{\widehat{\tau}_1}}
\newcommand{\whtauy}{{\widehat{\tau}_2}}
\newcommand{\whtauz}{{\widehat{\tau}_3}}
\def\ket#1{\mbox{$\displaystyle\vert\,#1\,\rangle$}}
\def\bra#1{\mbox{$\displaystyle\langle\,#1\,\vert$}}
\def\braket#1#2{\mbox{$\displaystyle\langle\,#1\,\vert\,#2\,\rangle$}}
\def\dangle#1{\frac{d\Omega_{#1}}{4\pi}}
\def\der#1#2{\mbox{$\displaystyle\frac{d #1}{d #2}$}}
\def\pder#1#2{\mbox{$\displaystyle\frac{\partial #1}{\partial #2}$}}
\newcommand{\He}{$^3$He}

\newcommand{\Hea}{$^3$He-A}
\def\whDelta{\widehat{\Delta}}
\def\nicefrac#1#2{\genfrac{}{}{}{1}{#1}{#2}}
\newcommand{\grad}{\mbox{\boldmath$\nabla$}}
\newcommand{\mfG}{\mathfrak{G}}
\newcommand{\G}{\mfG}
\newcommand{\whmfG}{\widehat{\G}}
\newcommand{\whGR}{\widehat{\mathfrak{G}}^{\text{R}}}

\newcommand{\whGM}{\widehat{\mathfrak{G}}^{\text{M}}}
\def\point#1#2{{\tt #1}_{\mbox{\footnotesize #2}}}
\def\vA{{\bf A}}
\def\vR{{\bf R}}
\def\vp{{\bf p}}
\def\vr{{\bf r}}
\def\vv{{\bf v}}
\def\vx{{\bf x}}

\def\vn{{\bf n}}
\def\vj{{\bf j}}
\def\vJ{{\bf J}}
\def\cA{{\mathcal A}}

\def\cD{{\mathcal D}}

\def\cF{{\mathcal F}}
\def\cG{{\mathcal G}}
\def\cH{{\mathcal H}}

\def\cN{{\mathcal N}}

\def\cP{{\mathcal P}}

\def\ns{\negthickspace}
\def\ms{\negmedspace}
\begin{document}
\title{Andreev bound states and \\ their signatures}
\author{J. A. Sauls}
\address{Fermilab-Northwestern Center for Applied Physics and Superconducting Technologies \& 
	 Department of Physics, Northwestern University, Evanston, IL, USA
}
\subject{
Condensed Matter Physics\\ 
Low Temperature Physics \\
Superconductivity\\ 
Quantum Fluids\\
Quantum Transport 
}
\corres{J. A. Sauls \\
\email{sauls@northwestern.edu}}
\begin{abstract}
Many of the properties of superconductors related to quantum coherence are revealed when the superconducting state is forced to vary in space - in response to an external magnetic field, a proximity contact, an interface to a ferromagnet, or to impurities embedded in the superconductor.
Amoung the earliest examples is \emph{Andreev reflection} of an electron into a retro-reflected hole at a normal-superconducting interface. In regions of strong inhomogeneity multiple Andreev reflection leads to the formation of sub-gap states, \emph{Andreev bound states}, with excitation energies below the superconducting gap.  These states play a central role in our understanding of inhomogeneous superconductors.  

The discoveries of unconventional superconductivity in many classes of materials, advances in fabrication of superconducting/ferromagnetic hybrids and nano-structures for confining superfluid $^3$He, combined with theoretical developments in topological quantum matter have dramatically expanded the significance of branch conversion scattering and Andreev bound state formation.
This collection of articles highlights developments in inhomogeneous superconductivity, unconventional superconductivity and topological phases of superfluid $^3$He, in which Andreev scattering and bound states underpin much of the physics of these systems. 

This article provides an introduction to the basic physics of Andreev scattering, bound-state formation and their signatures. The goal is both an introduction for interested readers who are not already experts in the field, and to highlight several examples in which branch conversion scattering and Andreev bound states provide unique signatures in the transport properties of superconductors.

This article is an introduction to the theme issue 'Andreev bound states'.
\end{abstract}
\maketitle
\section{Introduction} 

The BCS theory of superconductivity was a watershed in modern condensed matter physics \cite{bar57}. The key feature of the theory is \emph{pair condensation} - the macroscopic occupation of a bound state of Fermion pairs. The binding of Fermions into Cooper pairs typically leads to an energy gap in the Fermionic excitation spectrum, while condensation of Cooper pairs leads to the breaking of global $\point{U(1)}{\ns}$ gauge symmetry, the generator being particle number.
The latter implies that the Fermionic excitations are no longer charge eigenstates, but each is a coherent superposition of a normal-state particle and hole, e.g. $\gamma^{\dag} = u c^{\dag} + v c$, where $u$ and $v$ are the particle and hole amplitudes defining the Bogoliubov quasiparticles.
Charge conservation is maintained by an additional channel for charge transport via the coherent motion of the pair condensate.

Many of the remarkable properties of superconductors originate from the coherent superposition of particle and hole states - \emph{particle-hole coherence} - that defines the low-energy excitations of a superconductor, i.e. Bogolibov quasiparticles.
One feature of the normal metallic state that is preserved is that the Fermionic excitations of the superconducting state still come in two flavors: particle-like excitations with group velocity along the direction of the momentum, $\vv_{\vp}\cdot\vp > 0$, and hole-like excitations with reversed group velocity, $\bar\vv_{\vp}\cdot \vp < 0$.

\begin{figure}[t]
\begin{center}
\includegraphics[width=0.30\textwidth]{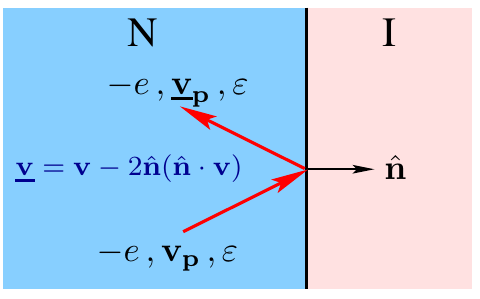}\qquad\includegraphics[width=0.30\textwidth]{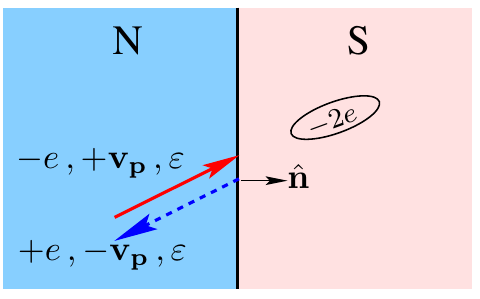}\qquad\includegraphics[width=0.30\textwidth]{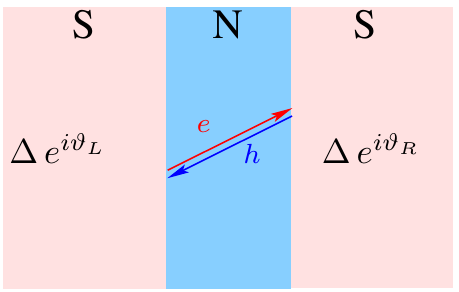}
\caption{Left: Specular reflection at an N-I boundary. 
         Center: Retro-reflection at an N-S boundary.
	 Right: Andreev-bound state confined within an S-N-S sandwich.}
\end{center}
\label{fig-specular-retro-reflections}
\end{figure}

The coherence amplitudes, $u$ and $v$, depend on the pair potential, $\Delta(\vr)$. Spatial variations of the pair potential lead to modifications of the coherence amplitudes, particularly to a novel scattering process, identified by A. F. Andreev, in which an incoming particle-like excitation has a finite probability to convert to an out-going hole-like excitation, a process called \emph{branch conversion} scattering, or Andreev scattering \cite{and64}.
When this process is combined with strong spatial variations or strong scattering, such as occurs in metal-superconductor proximity contacts, or cores of quantized vortices, or superconducing-ferromagnetic interfaces, \emph{multiple Andreev scattering} leads to the formation of \emph{Andreev bound states}, with sub-gap energies, that are localized near the region of strong spatial variations of $\Delta(\vr)$.
Indeed the simplest example of Andreev scattering is provided by the reflection of an electron in a normal (N) region by the pair potential of the superconducting region as shown in the center panel of Fig. \ref{fig-specular-retro-reflections}. 

The two branches of excitations of a normal metal, conduction electrons with charge $-e$, momentum $\vp$ and group velocity $\vv_{\vp} || \vp$, and holes - their anti-particle - with charge $+e$, momentum $\vp$ and reversed group velocity, $\bar\vv_{\vp} = - \vv_{\vp}$, are separately conserved in normal scattering processes by impurities, defects or surfaces. For example, an electron with momentum $\vp$, velocity $\vv_{\vp}$ and energy below the energy gap of an insulator is reflected by the insulating gap into a outgoing electron with momentum $\underline\vp=\vp-2\hat\vn(\hat\vn\cdot\vp)$, and group velocity $\vv_{\underline\vp}|| \underline\vp$, as shown in the left panel of Fig. \ref{fig-specular-retro-reflections}. This is specular reflection of electrons at a metal-insulator boundary.  

Andreev considered the reflection of electrons in a region of normal metal incident on a region of the same metal, but in the superconducting state, such as occurs in the intermediate state of a type I superconductor. For energies below the gap, $\varepsilon < \Delta$, the electron is forbidden to propagate into the superconducting region. However, upon reflection the outgoing excitation is a hole, i.e. scattering of an electron by the pair potential coverts and electron into a hole. Furthermore, the reflected hole is not specularly reflected, but \emph{retro-reflected} as illustrated in the center panel of Fig. \ref{fig-specular-retro-reflections}. Scattering that involves conversion of an electron into a hole, or vice-versa, is called \emph{branch conversion scattering}, or \emph{Andreev scattering}. Another generic feature of Andreev scattering is that there is very little change in momentum of the retro-reflected hole, $\delta p = 2\varepsilon/v_f \ll p_f$. Thus, both momentum and charge are transported across the normal-superconductor (N-S) interface. This is made possible by the transport of charge and momentum by Cooper pairs in the superconducting region.

\vspace*{-3mm}
\subsection{From Bogoliubov to Andreev}

The theory of inhomogeneous superconductors that goes beyond the limitations of the Ginzburg-Landau and London theories, and treats the effects of spatial variations of external fields and the self-consistent pair condensate on equal footing can be formulated starting from Gorkov's or Bogoliubov's mean-field theory of pairing \cite{gor58,bog58a}. The mean field pairing Hamiltonian,
\ber
\cH 
&=& 
\int d\vr\,\psi^{\dag}_{\alpha}(\vr)
	   \left[\frac{1}{2m^*}\left(\vp-\frac{e}{c}\vA(\vr)\right)^2-\mu\right]
           \psi_{\alpha}(\vr)
\nonumber\\
&-&
\frac{1}{2}
\int d\vr\int d\vr' 
\left\{
\psi^{\dag}_{\alpha}(\vr)
\psi^{\dag}_{\beta}(\vr')
\Delta_{\beta\alpha}(\vr',\vr)
+
\bar\Delta_{\alpha\beta}(\vr,\vr')
\psi_{\beta}(\vr')
\psi_{\alpha}(\vr)
\right\}
\,,
\eer
includes the kinetic energy of the normal-state Fermions,
where $\psi^{\dag}_{\alpha}(\vr)$ creates a normal-state conduction electron with spin projection, $\alpha \in \{\uparrow,\downarrow\}$ at $\vr$, and $\Delta_{\alpha\beta}(\vr,\vr')$ is the mean pairing energy, or \emph{pair potential} for the spin configuration $(\alpha,\beta)$ of the Cooper pairs, and $\bar\Delta_{\alpha\beta}(\vr,\vr') = -\Delta_{\alpha\beta}(\vr,\vr')^*$. 
Since $\cH$ is bilinear in the Fermion field operators it can be diagonalized by a canonical transformation, 
\be\label{eq-canonical_transformation}
\psi(\vr,\alpha)=\sum_n\,\left\{u_n(\vr,\alpha)\gamma_n+v_n(\vr,\alpha)\gamma_n^{\dag}\right\}
\,, 
\ee
where the $\left\{u_n(\vr,\alpha), v_n(\vr,\alpha)\right\}$ are the particle and hole amplitudes defining the Bogoliubov transformation to a new set of operators obeying Fermion anti-commutation relations: $\{\gamma_n\,,\,\gamma_{n'}^{\dag}\}=\delta_{n,n'}$ and $\{\gamma_n\,,\,\gamma_{n'}\}= 0$, which diagonalizes the mean-field Hamiltonian, $\cH = E_{s} + \sum_n\,\varepsilon_n\,\gamma_{n}^{\dag}\gamma_{n}$, where the sum is over all positive energy states. Occupying the negative energy states generates the BCS ground state, and accounts for the ground-state energy, $E_s$.
The mode sum in Eq. \ref{eq-canonical_transformation} runs over a complete set of orthonormal states $\{u_n,v_n|\forall n\}$, where the quantum numbers $\{n\}$ depend on the geometry and symmetry of the confining potential, and inhomogeneities generated by the external field and pair potential. The required amplitudes, $u_n(\vr,\alpha)$ and $v_n(\vr,\alpha)$, obey the Bogoliubov equations \cite{kur90},\footnote{For unconventional superconductors the pair potential depends two spatial coordinates reflecting the internal orbital state of the Cooper pairs that form the condensate. Thus, it is most convient to transform to the center or mass and relative coordinates, $\vR = (\vr+\vr')/2$ and $\vx=\vr-\vr'$. Since the radial extent of the Cooper pair wavefunction is typically large compared to the Fermi wavelength, $\xi \gg \hbar/p_f$, the relative-momentum-space wavefunction depends on momenta that are concentrated near the Fermi surface. Thus, we Fourier transform the relative coordinate and express the pair potential as $\Delta_{\alpha\beta}(\vR,\vp)$.}

\ber
\varepsilon u(\vr,\alpha) 
&=& 
+ \left(\frac{1}{2m^*}\left(\vp-\frac{e}{c}\vA(\vr)\right)^2-\mu\right)
u(\vr,\alpha) + \Delta_{\alpha\beta}(\vr,\vp)\,v(\vr,\beta)
\,,
\label{eq-Bogoliubov-u}
\\
\varepsilon v(\vr,\alpha) 
&=& 
-\left(\frac{1}{2m^*}\left(\vp-\frac{e}{c}\vA(\vr)\right)^2-\mu\right)
v(\vr,\alpha) + \Delta^{\dag}_{\alpha\beta}(\vr,\vp)\,u(\vr,\beta)
\,,
\label{eq-Bogoliubov-v}
\eer
where $\vp \rightarrow \frac{\hbar}{i}\grad$. These equations determine the particle and hole amplitudes that define the Bogoliubov quasiparticle excitations of an inhomogeneous superconductor described by the spin-dependent mean-field pair potential, $\Delta_{\alpha\beta}(\vr,\vp)$.

For a homogeneous superconductor in zero magnetic field the solutions are momentum eigenstates, with amplitudes $(u_{\vp\alpha},v_{\vp\alpha})$ determined by the $4\times 4$ matrix eigenvalue equation, $\ket{\varphi}=\mbox{col}\left(u_{\vp\uparrow}\,\,u_{\vp\downarrow}\,\,v_{\vp\uparrow}\,\,v_{\vp\downarrow}\right)$ 
and 
\be\label{eq-Bogoliubov2}
\widehat{\cH}_{\text{B}}\ket{\varphi} 
= 
\varepsilon
\ket{\varphi}
\leadsto
\begin{pmatrix} 
\xi_{\vp} \hat{1} & \hat\Delta(\vp) 
\\
\hat\Delta^{\dag}(\vp) & -\xi_{\vp}\hat{1}
\end{pmatrix}
\begin{pmatrix}
\hat{u}_{\vp} 
\\
\hat{v}_{\vp} 
\end{pmatrix}
=
\varepsilon
\begin{pmatrix}
\hat{u}_{\vp} 
\\
\hat{v}_{\vp} 
\end{pmatrix}
\,,
\ee
where $\hat{1}$ is the unit matrix in spin space, $\hat\Delta(\vp)$ is the $2\times 2$ spin matrix order parameter, $\xi_{\vp}=|\vp|^2/2 m^* - \mu$ is the normal-state excitation energy, and 
$\hat{u}_{\vp}$ 
($\hat{v}_{\vp}$)
is the two-component particle (hole) spinor.

In what follows I discuss Andreev scattering, bound states and their signatures for conventional superconductors. Andreev scattering and bound states in d-wave superconductors, and unconventional spin-triplet superconductors, including superfluid \He, and superconducting-ferromagnetic hybrids are discussed in several articles in this volume. These superconductors are generally derived from parent states that are invariant under space inversion. Furthermore, in most cases of interest the superconducting ground states are either spin-singlet or \emph{unitary} spin-triplet states. The latter break spin-rotation symmetry, but the Cooper pairs have no net spin polarization along any direction in spin space.
Inversion symmetry implies that the pairing interaction responsible the Cooper instability separates into even- and odd-parity sectors, and thus the Cooper pairs have definite parity, i.e. $\hat\Delta(-\vp) = \pm\hat\Delta(\vp)$. Combined with the anti-symmetry of the condensate amplitude under Fermion exchange, 
$\Delta_{\alpha\beta}(\vp) = -\Delta_{\beta\alpha}(-\vp)$, the possible superconducting classes divide into
even-parity, spin-singlet states and odd-parity, spin-triplet states with 
\be\label{eq-Singlet-Triplet}
\hat\Delta(\vp) = 
\Bigg\{
\begin{array}{ll}
\Delta(\vp)\,i\sigma_y 
& 
\quad \mbox{even parity, spin}\,$S=0$ 
\,,
\\
\vec{\Delta}(\vp)\,\cdot\,i\vec{\sigma}\sigma_y 
&
\quad \mbox{odd parity, spin}\,$S=1$
\,,
\end{array}
\ee
expressed in terms of the anti-symmetric Pauli matrix, $i\sigma_y$, for spin-singlet pairing, and the three symmetric Pauli matrices, $i\sigma_i\sigma_y$ for $i=x,y,z$, for spin-triplet pairing. Thus, $\Delta(\vp)$ ($\vec\Delta(\vp)$) is a complex scalar (vector) under spin rotations.
%
\begin{figure}[t]
\centering
\includegraphics[width=0.65\textwidth]{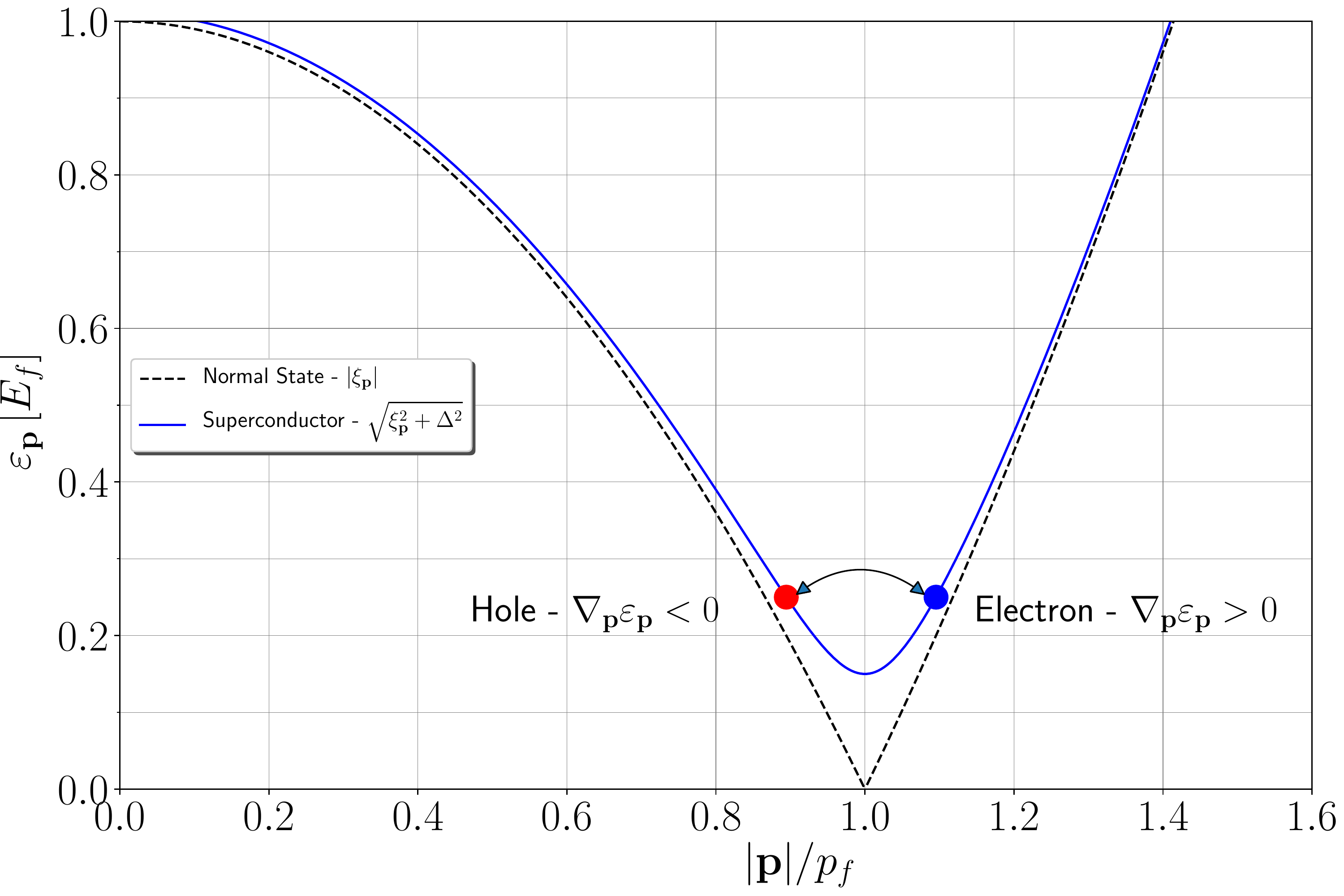}
\caption{Dispersion relations for the normal state (black dashed line) and superconducting state (solid blue line). Branch conversion scattering is indicated between an electron-like (right blue dot) excitation and a hole-like (left red dot) excitation of the same energy. For illustrative purposes the excitation gap was set at $\Delta = 0.15\,E_f$.}
\label{fig-electron-hole_dispersion}
\end{figure}
%
The general class of ``unitary'' states satisfy the condition, 
\be\label{eq-unitarity}
\hat\Delta(\vp)\hat\Delta^{\dag}(\vp) = |\Delta(\vp)|^2\,\hat{1}
\,,
\ee
which for spin-triplet pairing implies that the Cooper pair spin polarzation, $\vec{S}_{\text{pair}} \propto i\vec\Delta(\vp)\times\vec\Delta(\vp)^* \equiv 0$ at each point $\vp$. The superfluid phases of \He\ belong to the unitary spin-triplet class.
The resulting excitation spectrum follows immediately from Eqs. \ref{eq-Bogoliubov2} and \ref{eq-unitarity}; there branches of positive and negative energy states, $\varepsilon^{(\pm)}_{\vp} = \pm\sqrt{\xi_{\vp}^2 + |\Delta(\vp)|^2}$. The negative energy states are filled and define the particle-hole coherent ground state. 

The positive energy states are the Bogoliubov excitations of the superconductor. For momenta near the Fermi surface $|\vp|\sim p_f$, the normal-state excitation spectrum is linear in the momentum, $\varepsilon_{\vp} = |\xi_{\vp}| \approx v_f| p - p_f|$, and there are two branches: electrons with group velocity $v_{\vp} = v_f\hat{p}$, and holes with $\bar{v}_{\vp} = -v_f\hat{p}$. 
The corresponding electron-like and hole-like excitations of the superconductor have group velocities,
\be
v_{\hat\vp}^{(\pm)}(\varepsilon) = \pm v_f\,\frac{\sqrt{\varepsilon^2-|\Delta(\hat\vp)|^2}}{\varepsilon}
\,,\quad \varepsilon\ge|\Delta(\hat\vp)|
\,.
\ee
Thus, the speed of Bogoliubov excitations becomes vanishingly small as the energy approaches the gap edge. At energies above the gap variations of the pair potentical provide a mechanism for Andreev scattering between electron-like and hole-like excitaitons at the same energy, as indicated in Fig. \ref{fig-electron-hole_dispersion}

In his treatment of heat transport in the intermediate state Andreev provided an important simplification of Bogoliubov's equations for the Fermionic excitations of an inhomogeneous superconductor, which plays an central role in theoretical developments in inhomogeneous superconductors and superconductivity under non-equilibrium conditions \cite{eil68,lar69,ser83,she85,mil88}.
The key observation is that the superconducting correlation length, $\xi_0=\frac{\hbar v_f}{2\pi k_{\text{B}} T_c}$, which is  characteristic length scale for spatial variations of the pair potential, is typically much longer than the Fermi wavelength, $\lambda_f = \hbar/p_f$. 
Thus, for electronic excitations of the superconducting state this separation of scales leads to fast spatial variations on the scale of $\lambda_f$, modulated by slow spatial variations resulting from the pair potential. Furthermore, the excitation gap is small compared to the Fermi energy, $|\Delta(\vp)| \ll E_f$.
Thus, we look for solutions of Eqs. \ref{eq-Bogoliubov-u}-\ref{eq-Bogoliubov-v} by factoring the ``fast oscillations'' at the Fermi wavelength, $\ket{\varphi(\vr)} = e^{i\vp_f\cdot\vr/\hbar}\,\times\,\ket{\Psi_{\vp}(\vr)}$, and retaining the leading order terms in $\hbar/p_f\xi_0$ and $|\Delta(\vp)|/E_f$. The latter also implies that the pair potential can be evaluated for momenta, $\vp\approx p_f\hat\vp$.
The slow spatial variations of the Bogoliubov spinors is governed by Andreev's equation \cite{and64},
\be
\left[\varepsilon\whtauz - \whDelta(\vp,\vr)\right]\ket{\Psi_{\vp}(\vr)} 
+ i\hbar\vv_{\vp}\cdot
\left(\grad - i\nicefrac{2e}{\hbar c}\vA(\vr)\right)\ket{\Psi_{\vp}(\vr)}
= 0
\,,
\label{eq-Andreev_Equation}
\ee
where $\vp=p_f\hat\vp$, $\vv_{\vp} = v_f\hat\vp$, and the pair potential has been redefined by, $\whtauz\whDelta\rightarrow\whDelta$. Particle-hole coherence is encoded in the Andreev spinor $\ket{\Psi_{\vp}(\vr)} = \mbox{col}\left(u_{\vp\uparrow}(\vr)\,\,u_{\vp\downarrow}(\vr) \,\,v_{\vp\uparrow}(\vr) \,\,v_{\vp\downarrow}(\vr)\right)$, and necessarily on the longer wavelength scale set by the pair potential.

Andreev's equation is a first-order differential equation for the evolution of the particle-hole states along ``trajectories'' defined by a point $\hat\vp$ on the Fermi surface. For a normal metal in zero magnetic field Andreev's equation reduces to decoupled characteristic equations for electron and hole states with the replacement $\varepsilon\rightarrow i\hbar\partial_t$,
\ber
\left(\partial_t + \vv_{\vp}\cdot\grad\right)\,u_{\vp} = 0  
\,,\quad
\left(\partial_t - \vv_{\vp}\cdot\grad\right)\,v_{\vp} = 0  
\,.
\eer
Thus, for each point on the Fermi surface the electron ($e$) and hole ($h$) spinors are given by
\be
\ket{\Psi^{(e)}_{\vp,s}} = \begin{pmatrix} \chi_s \\ 0 \end{pmatrix}\,
e^{-i\varepsilon t/\hbar}\,e^{i\varepsilon\,\hat\vp\cdot\vr/\hbar v_f}
\,,
\qquad
\ket{\Psi^{(h)}_{\vp,s}} = \begin{pmatrix} 0 \\ (-i\sigma_y)\chi_s \end{pmatrix}\,
e^{-i\varepsilon t/\hbar}\,e^{-i\varepsilon\,\hat\vp\cdot\vr/\hbar v_f}
\,,
\ee
where $\chi_s$ is the two-component spinor. Note the convention: $(-i\sigma_y)$ rotates the spinor $\chi_s$ by $180^{o}$ about the $y$-axis. The energy-momentum relation of the electron (hole) solution is $\vp^{({e\atop h})}=(p_f \pm \varepsilon/v_f)\hat\vp$, and the group velocity is 
$\vv^{({e\atop h})}_{\vp}=\pm v_f\hat\vp$. 

For a homogeneous, conventional spin-singlet superconductor in zero magnetic fielld the pair potential is independent of $\vp$ and $\vr$. The eigenstates of a homogeneous superconductor are solutions of the form, $\ket{\Psi^{\lambda}_{\vp,s}} = e^{i\lambda\,\hat\vp\cdot\vr/\hbar v_f}\,\ket{\lambda}$, where $\lambda$ is an eigenvalue and $\ket{\lambda}=\mbox{col}(u_{\lambda}\,\,v_{\lambda})$ is an eigenvector of 
\be
\begin{pmatrix} 
\varepsilon\hat{1} & -\Delta\,(i\sigma_y)
\\ 
\Delta^*\,(-i\sigma_y) & -\varepsilon\hat{1}
\end{pmatrix}
\begin{pmatrix} 
	u_{\lambda} \\ v_{\lambda}
\end{pmatrix}
=
\lambda
\begin{pmatrix} 
	u_{\lambda} \\ v_{\lambda}
\end{pmatrix}
\,.
\ee
For $\varepsilon>\Delta$ there are two eigenvalues, $\lambda_{\pm}=\pm\lambda(\varepsilon)$, where $\lambda(\varepsilon)\equiv\sqrt{\varepsilon^2-|\Delta|^2}$, each with a two-fold spin degeneracy. The corresponding eigenvectors are the electron-like and hole-like Bogoliubov excitaitons of the homogeneous bulk superconductor,
The electron-like solution is 
\be\label{eq-Andreev-electron}
\ket{\Psi^{(e)}_{\vp,s}} =\frac{1}{\sqrt{2\varepsilon(\varepsilon+\lambda)}}
\begin{pmatrix} 
(\varepsilon+\lambda)\,\chi_s 
\\ 
\Delta^*\,(-i\sigma_y)\,\chi_s
\end{pmatrix}
\,
e^{i\lambda\hat\vp\cdot\vr/\hbar v_f}
\xrightarrow[]{\varepsilon\gg |\Delta|}
\begin{pmatrix} 
\chi_s 
\\ 
0
\end{pmatrix}
\,
e^{i\varepsilon\hat\vp\cdot\vr/\hbar v_f}
\,,
\ee
and the hole-like solution is 
\be\label{eq-Andreev-hole}
\ket{\Psi^{(h)}_{\vp,s}} =\frac{1}{\sqrt{2\varepsilon(\varepsilon+\lambda)}}
\begin{pmatrix} 
-\Delta\,\chi_s
\\ 
(\varepsilon+\lambda)\,(-i\sigma_y)\,\chi_s 
\end{pmatrix}
\,
e^{-i\lambda\hat\vp\cdot\vr/\hbar v_f}
\xrightarrow[]{\varepsilon\gg |\Delta|}
\begin{pmatrix} 
0
\\ 
(-i\sigma_y)\chi_s 
\end{pmatrix}
\,
e^{-i\varepsilon\hat\vp\cdot\vr/\hbar v_f}
\,.
\ee
All four states are mutually orthonormal, i.e. $\braket{\Psi^{t}_{\vp,s}}{\Psi^{t'}_{\vp,s'}} = \delta_{t,t'}\delta_{s,s'}$, for $t,t'\in\{e,h\}$ and $s,s'\in\{\uparrow,\downarrow\}$. Thus, an electron-like excitation with spin $s=\uparrow$ is a superposition of a normal conduction electron with spin $s=\uparrow$ and a hole with spin $s=\downarrow$.

Equations \ref{eq-Andreev_Equation} also have solutions for sub-gap energies, $\varepsilon < |\Delta|$. However, the eigenvalues are pure imaginary, $\lambda_{\pm}=\mp\Lambda(\varepsilon)$ with $\Lambda(\varepsilon) \equiv\sqrt{|\Delta|^2-\varepsilon^2}$, 
\be\label{eq-Andreev-Evanescent}
\ket{\Psi^{(\pm)}_{\vp,s}} = B_{\pm}
\begin{pmatrix} 
(\varepsilon\mp i\Lambda)\,\chi_s 
\\ 
\Delta^*\,(\mp i\sigma_y)\,\chi_s
\end{pmatrix}
\,
e^{\pm\Lambda\hat\vp\cdot\vr/\hbar v_f}
\,.
\ee
These solutions explode exponentially along the trajectory coordinate, $x=\hat{p}\cdot\vr$, and are non-normalizeable, and thus un-physical solutions for a homogeneous bulk superconductor, i.e. there are no sub-gap states of a homogeneous s-wave superconductor.
However, these exploding/decaying solutions are allowed for finite geometries, such as N-S interfaces, or regions where the pair potential varies sharply in space, such as normal-metal inclusions in a superconductor, or the cores of vortices \cite{rai96,sau09}.

What is clear from the solutions to Andreev equations is that the particle-hole degree of freedom of excitations of the supercondutor is a two-component iso-spin analogous to the spin degree of freedom. Just as the spin state of a quasiparticle evolves smoothly, ``rotates'', in the presence of a a slowly varying Zeeman potential, or undergoes a spin-flip transition induced by a rapidly varying field, the particle-hole isospin can rotate slowly in particle-hole space under the action of a slowly varying pair potential, or undergo a branch conversion (iso-spin-flip) when acted upon by a rapidly varying pair potential. The latter case is illustrated by Andreev reflection at an N-S interface (center panel of Fig. \ref{fig-specular-retro-reflections}).

\subsubsection{Andreev Reflection at N-S Boundaries}

For an electron incident from the normal metallic (N) region with energy below the superconducting (S) gap there is no propagating solution in the S region. However, an incident electron penetrates a short distance into the S region as an evanescent coherent particle-hole excitation decaying into the S region. The discontinuity of the pair potential at the N-S requires a retro-reflected hole. Thus, the scattering solution that conserves spin for $x<0$ is
\be\label{eq-N-region}
\ket{\Psi^{<}_{\vp,s}(x)} 
=
\begin{pmatrix} \chi_s \\ 0 \end{pmatrix}\,e^{+i\varepsilon x/\hbar v_f} 
+r_{\text{A}}\,\begin{pmatrix} 0 \\ (-i\sigma_y)\chi_s \end{pmatrix}\,e^{-i\varepsilon x/\hbar v_f} 
\,,
\ee
where $r_A$ is the amplitude of the reflected hole. For $x>0$ only an evanescent solution decaying into the S region is physical,
\be\label{eq-S-region-evanescent}
\ket{\Psi^{>}_{\vp,s}(x)}
= 
\frac{1}{\varepsilon +i\Lambda}
\begin{pmatrix} 
(\varepsilon +i\Lambda)\,\chi_s 
\\ 
\Delta^*\,(i\sigma_y)\chi_s
\end{pmatrix}\,
e^{-\Lambda x/\hbar v_f} 
\,.
\ee
The boundary condition at the N-S interface yields the Andreev reflection amplitude,
\be\label{eq-Andreev_reflection_sub-gap}
r_{\text{A}} = -\frac{\Delta^{*}}{\varepsilon+i\sqrt{|\Delta|^2 - \varepsilon^2}}
\,,\quad \varepsilon \le |\Delta|
\,.
\ee
Note that a retro-reflected hole is certain for $\varepsilon<|\Delta|$, i.e. the reflection probability is $|r_{\text{A}}(\varepsilon)|^2 \equiv 1$.

Andreev reflection also occurs for electrons incident with energies $\varepsilon>|\Delta|$, ``above-gap reflection''. Andreev scattering for energies above the gap is a purely quantum mechanical result. For $\varepsilon>|\Delta|$ the scattering solution for $x>0$ is a particle-like excitation propagating into the S region. There is no hole-like excitation propagating towards the N-S interface from the S region,
\be\label{eq-S-region-propagating}
\ket{\Psi^{>}_{\vp,s}(x)}
= 
\frac{1}{\varepsilon+\lambda}
\begin{pmatrix} 
(\varepsilon+\lambda)\,\chi_s 
\\ 
\Delta^*\,(-i\sigma_y)\,\chi_s
\end{pmatrix}\,
e^{+i\lambda x/\hbar v_f} 
\,,
\quad \varepsilon\ge |\Delta|
\,.
\ee
The resulting Andreev reflection amplitude is, 
\be
r_{\text{A}} = \frac{\Delta^*}{\varepsilon+\sqrt{\varepsilon^2-|\Delta|^2}}
\,.
\ee
Thus, the \emph{probabilty} for retro-reflection is
\be
|r_{\text{A}}|^2 
= 
\Bigg\{
\begin{tabular}{cl} 
$1$
& 
$\,,\,\varepsilon\le |\Delta|$
\\
$\displaystyle{\frac{|\Delta|^2}{(\varepsilon+\sqrt{\varepsilon^2-|\Delta|^2})^2}}$
& 
$\,,\,\varepsilon > |\Delta|$\,\,.
\end{tabular}
\ee

\subsubsection{N-S Boundary: Heat Conductance}

For a particular trajectory $\vp$ and energy $\varepsilon<|\Delta|$ the charge transported across the N-S interface from the N region is 
\ber
\vj_{e} = (+e)\vv^{(e)}_{\vp} + (-e)\vv^{(h)}_{\vp}*|r_{\text{A}}|^2
        = 2ev_f\,\hat\vp\,,\quad \varepsilon\le|\Delta|
\,. 
\eer
%
Since there are no quasiparticle states below the gap of the superconductor, these sub-gap currents are transported by the condensate of Cooper pairs in the S region.

Electron and hole excitations in the N region also transport energy. For the states defined by the trajectory $\vp$ and excitation energies $\varepsilon \le |\Delta|$ the heat current is $\vj_{q} = \varepsilon\vv^{(e)}_{\vp} + \varepsilon\vv^{(h)}_{\vp}*|r_{\text{A}}(\varepsilon)|^2 \equiv 0$, which is expected and necessary as there are no states to transport heat energy in the S region, and the Cooper pair condensate carries zero entropy. However, for excitation energies $\varepsilon>|\Delta|$ there is a net heat current in the N region that is transported by states above the gap in the S region, leading to a finite thermal conductance for the N-S boundary. 
In particular, for a thermal bias $\delta T$ across the N-S regions the number of excess carriers at energy $\varepsilon$ in the hotter N region is $\delta f(\varepsilon) = (\partial f/\partial T) \delta T$, where $f(\varepsilon) = 1/(1+e^{\varepsilon/T})$ is the Fermi distribution. Thus, the heat current transported from the N region to across the N-S interface is
\ber
\vJ_{q} = \cA\,\int_{+}\frac{d\Omega_{\hat\vp}}{4\pi}\,2N_f\,
               \int_{0}^{\infty}\,d\varepsilon\,
		\delta f(\varepsilon)
               (\varepsilon\vv_{\vp})\,
               \left[1 - |r_{\text{A}}(\varepsilon)|^2\right]
	= \kappa\,\delta T\,\hat\vn
\,,
\eer
where the integration is over half the Fermi surface corresponding to excitations with $\hat\vp\cdot\hat\vn \ge 0$ with $\hat\vn$ directed into the S region normal to the N-S interface (Fig. \ref{fig-specular-retro-reflections}), and $\cA$ is the area of the interface. The resulting thermal conductance includes the additional 
suppression 
of the heat current resulting from the finite probability of Andreev reflection of excitations with energies 
above the gap, 
\be
\kappa = \nicefrac{1}{2}N_f\,v_f\,\cA\,
               \int_{|\Delta|}^{\infty}\,d\varepsilon\,
               \frac{\varepsilon^2/4T^2}{\cosh^2(\varepsilon/2T)}\,
               \left[1 - |r_{\text{A}}(\varepsilon)|^2\right]
\,.
\label{eq-conductance-AR}
\ee
For an N-N contact of area $\cA$ and temperature drop $\delta T$ we obtain the heat conductance of a Sharvin contact with two normal metallic leads, $\kappa_{\text{N}} = \cA\,\frac{\pi^2}{12}\,N_f\,v_f\,T$. Similarly, for the S-S contact we obtain the Sharvin heat conductance for two superconducting leads, which is Eq. \ref{eq-conductance-AR} without Andreev reflection. 

\begin{figure}[t]
\centering
\includegraphics[width=0.65\textwidth]{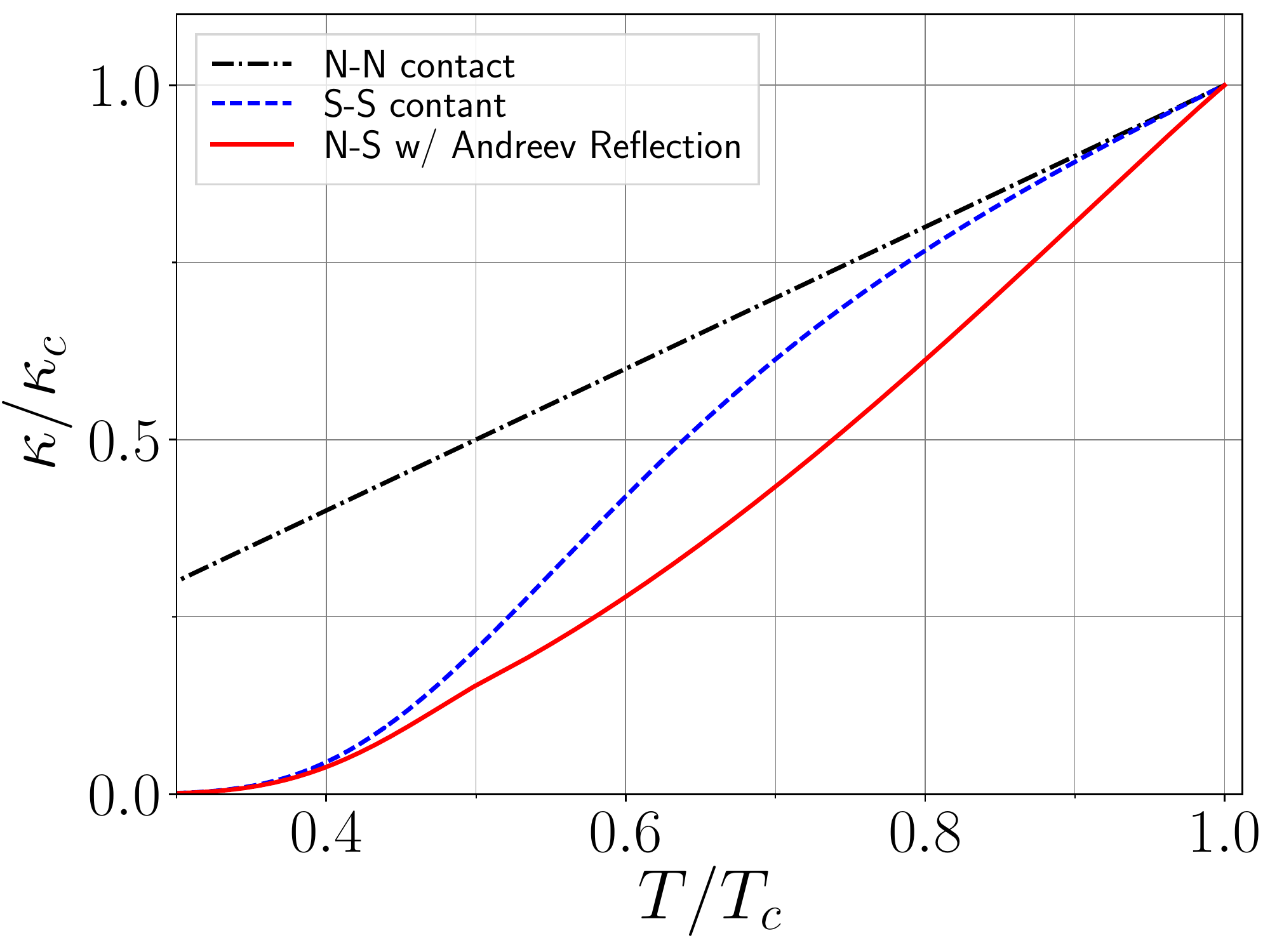}
\caption{
Thermal conductance of an N-N (black dot-dashes), S-S (blue dashes) and N-S (red line) contact. Andreev reflection of quasiparticles for energies \emph{above} the gap for an N-S contact surpresses the conductance below that of an S-S contact.  
         }
\label{fig-conductance-AR}
\end{figure}

Figure \ref{fig-conductance-AR} shows the thermal conductance of N-N, S-S and N-S contacts.
The temperature dependence of the S-S contact is the same as the thermal conductivity of a bulk BCS superconductor with non-magnetic impurities, which reflects the suppression of the number of excitations that can transport heat due to the gap in the spectrum. The conductance of the N-S contact shows the additional suppression of the heat transport resulting from Andreev reflection of excitations for $\varepsilon\ge|\Delta|$.
This result is the basis for the general view that branch conversion scattering suppresses heat transport. 
In Sec. \ref{sec_Point-Contact-Heat-Transport} I discuss branch conversion scattering and charge and heat transport in superconducting point-contact Josephson junctions. Both charge and heat currents depend on the phase difference of between the order parameters of the two superconducting leads. Under certain conditions of phase bias and temperature bias branch-conversion scattering leads to strong \emph{enhancement} of heat transport below $T_c$. The origin of this effect is \emph{resonant transmission} of quasiparticles that results from a shallow Andreev bound state below the continuum edge.

\subsection{Bound-State Formation: S-N-S Weak Links}

For spatial variations of the order parameter that lead to quantum interference of multiple Andreev reflections bound states of electron-like and hole-like excitations, i.e. Andreev bound states, form \cite{and66a}. The simplest geometry for Andreev bound state formation is the S-N-S structure shown in the right panel of Fig. \ref{fig-specular-retro-reflections}. The pair potential of the left (L) nad right (R) lead is $\Delta_{L,R} = |\Delta|\,e^{i\vartheta_{L,R}}$. Thus, for energies below the excitation gap in either lead only exponetially decaying solutions are allowed in the left and right leads,
\be
\ket{\Psi_{L}} = A_{L}\,\begin{pmatrix}	
(\varepsilon-i\Lambda)\chi_{s}
\\ 
\Delta_L^*\,(-i\sigma_y)\chi_{s} 
\end{pmatrix}
\,e^{+\Lambda (x+d/2)/\hbar v_f}
\,,\qquad
\ket{\Psi_{R}} = A_{R}\,\begin{pmatrix}	
(\varepsilon+i\Lambda)\chi_{s}
\\ 
\Delta_R^*(i\sigma_y)\chi_{s}
\end{pmatrix}
\,e^{-\Lambda (x-d/2)/\hbar v_f}
\ee
and counter-propagating solutions within the N region,
\be
\ket{\Psi_{N}} = 
B^{<}_{N}\,
\begin{pmatrix}	
\chi_{s}
\\ 
0
\end{pmatrix}
\,e^{+i\varepsilon x/\hbar v_f}
\,\,+\,\, 
B^{>}_{N}\,
\begin{pmatrix}	
0
\\ 
(-i\sigma_y)\chi_{s}
\end{pmatrix}
\,e^{-i\varepsilon x/\hbar v_f}
\,.
\ee
Matching the solutions at $x = -d/2$ and $x = +d/2$ generates the eigenvalue equation for the 
spectrum of Andreev bound states confined by the S-N-S structure,
\ber
\frac{\varepsilon}{|\Delta|} 
= (-1)^{m}\cos\left(\bar{d}\,\frac{\varepsilon}{|\Delta|} + \vartheta/2\right)
\,,\quad m = 0,1
\,,
\eer
where 
$\bar{d}=d/\hat\vp\cdot\hat\vn\,\xi_{\mbox{\tiny $\Delta$}}$ is the effective width of the N region for the trajectory $\vp$, in units of the coherence length, $\xi_{\mbox{\tiny $\Delta$}} = \hbar v_f/|\Delta|$, and $\vartheta \equiv\vartheta_R-\vartheta_L$ is the phase difference between the two S regions.
For $\bar{d} \gg 1$ the number of bound states scales with the effective thickness, 
$N\approx\mbox{Integer}[\bar{d}/\pi]$, while for $\bar{d} < 1$ there is at least one bound state. 
For zero phase bias, $\vartheta = 0$, the bound state energy lies near the gap edge, 
$\varepsilon_{\text{abs}} \approx |\Delta| (1 - \nicefrac{1}{2}\,\bar{d}^2)$. 

\begin{figure}[t]
\centering
\begin{tabular}{lll}
\raisebox{0.5\totalheight}{\hspace*{-5mm}
\includegraphics[width=0.60\textwidth]{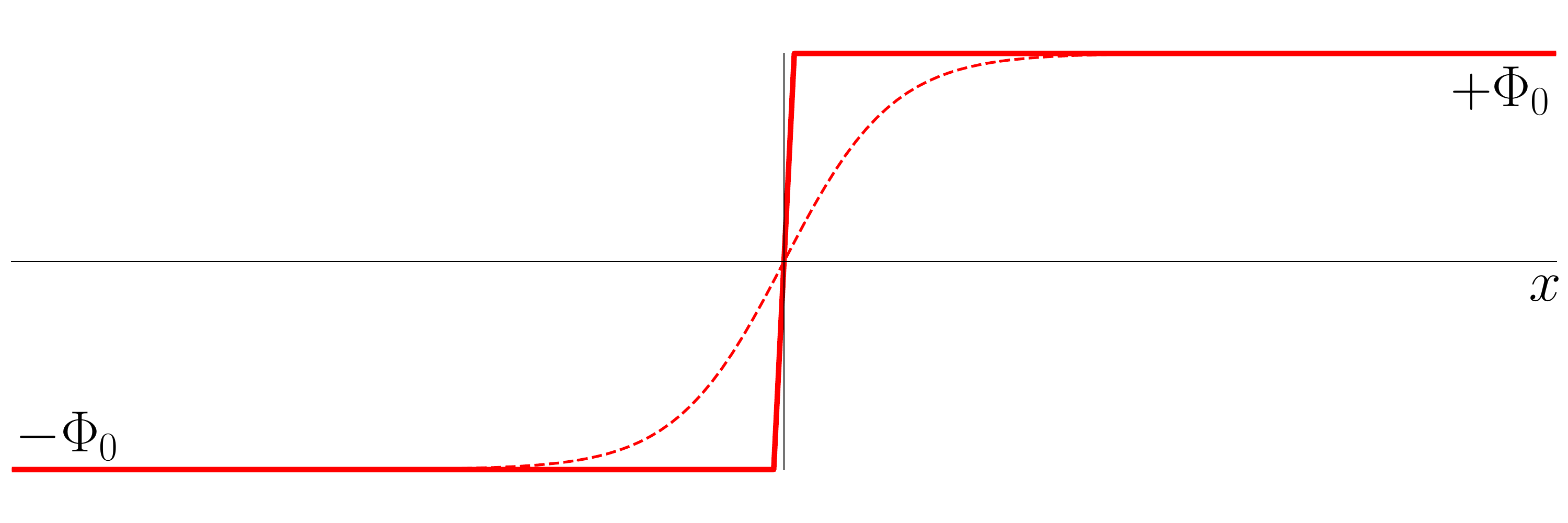}}
&
\qquad
&
\includegraphics[width=0.275\textwidth]{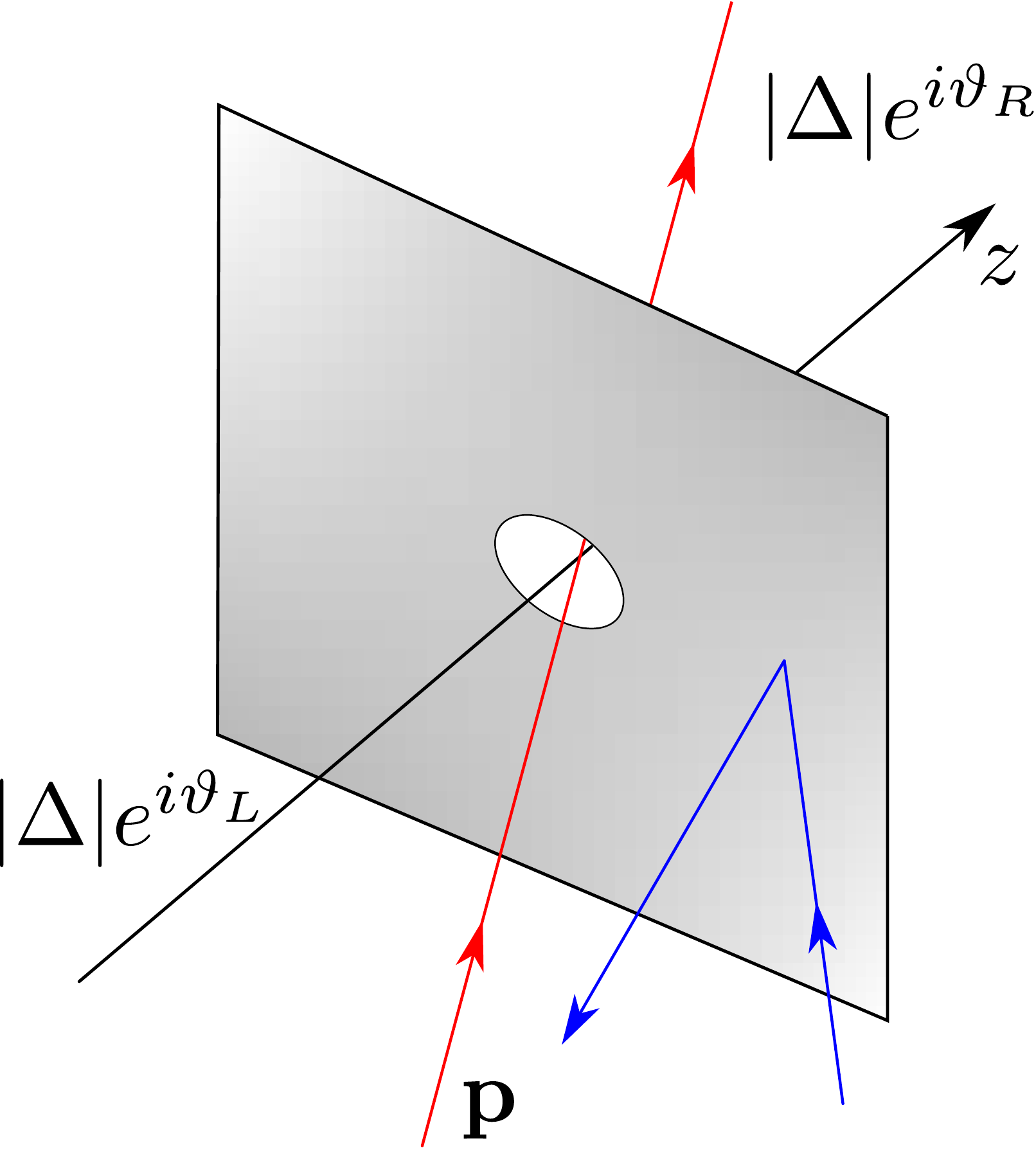}
\end{tabular}
\caption{Left: Domain wall: $\pi$ phase change of the superconducting order parameter. Right: Josephson point contact with radius $a\ll\xi_{\text{$\Delta$}}$. Current is transported only via trajectories passing through the aperature. Right: Branches of Andreev bound states of a S-S weak link dispersing as a function of the phase bias, $\vartheta$.
}
\label{fig_Josephson-Pinhole}, 
\end{figure}

\subsection{Connections: Andreev, Dirac and Jackiw-Rebbi Zero Modes}

However, the situation is different if the phase bias is tuned to $\vartheta = \pi$. In this case there is always a bound state with zero energy, even in the limit $\bar{d} \rightarrow 0$, i.e. the limit in which the order parameter changes sign across the S-S contact as shown in Fig. 
\ref{fig_Josephson-Pinhole}. 
The zero-energy bound state for the case of an S-S contact with a $\pi$ phase change is a known in many other physical contexts \cite{su79,hu94}, and is a realization of the Jackiw-Rebbi zero-energy bound state of relativistic fermions (Dirac fermions) coupled to a real scalar field with a domain wall separating two degenerate vacua,
\be
\left[i\hbar(\partial_t + c\vec{\alpha}\cdot\grad) + \beta\,g\Phi\right]\ket{\psi} = 0
\,,
\ee 
where $\ket{\psi} = \mbox{col}(\psi_1,\psi_2,\psi_3,\psi_4)$ is the four-component Dirac spinor, $c\vec\alpha$ is the light speed velocity operator, $\beta$ is 4-component mass operator  where $\Phi(x\rightarrow \pm\infty) = \pm \Phi_0$ is the scalar field with a domain wall - a ''kink'' - separating degenerate vacuum states with $\mp\Phi_0$, and $g$ is the coupling that generates the mass of the Dirac fermions \cite{jac76}. 

Thus, there is a strong connection to Bogoliubov fermions governed by Andreev's equation in which the pair potential is the counterpart of the scalar field. In both cases the excitations are described by four-component spinors satisfying a field equation that is linear in 
space-time derivatives. The pair potenial in Andreev's equation plays the role of the scalar field of the Jackiw-Rebbi theory.
The zero-energy bound state is topologically protected; i.e. it is robust to spatial variations of the $\Phi$ field, or the pair potential, so long as asymptotically the field describes degenerate vaccum states with Atiyah-Singer indices $\sgn\,{\Phi(x\rightarrow\pm\infty)}=\mp$ \cite{ati75}.
A key difference between Jackiw-Rebbi theory and Andreev theory is the pair potential is a complex scalar field, with a continuous degeneracy of the vacuum manifold. Thus, the bound-state energy disperses with phase difference, $\vartheta$, between the two S regions. This situation is realized by Josephson point contacts.

\subsection{Sharvin Contact - Andreev Bound States and Josephson Currents}\label{sec-Josephson-Pinhole-Conductance} 

The Josephson effect in superconductors connected by a small aperature, a ``pinhole'' in an otherwise insulating barrier separating the two superconductors, provides a textbook study of Andreev bound-state formation at the interface between two degenerate superconducting vacua.
The d.c. Josephson effect in pinhole junctions with ballistic electron transport was considered by Kulik and Omelyanchuck (K-O) \cite{kul77}. They calculated the dependence of the supercurrent flowing through the point contact as a function of the phase bais across the point contact, i.e. the current-phase relation, $I_s(\vartheta)$. Their result reduces to the standard Josephson current-phase relation, $I_s = I_c\sin\vartheta$, in the Ginzburg-Landau region, $|T-T_c| \ll T_c$, but deviates significantly from $I_s\propto\sin\vartheta$ at low temperatures, developing a discontinuity at $\vartheta = \pi/2$ at $T=0$, with a divergence of the slope 
$dI_s/d\vartheta\vert_{\vartheta=\pi}\sim -(T_c/T)I_c(0)$ for \quad $T\ll T_c$.
Although K-O did not explicitly discuss the role of the sub-gap Andreev bound state spectrum in their original papers, the anomalies in the current-phase relation of the Josephson point contact are reflections of the current transported by the spectrum of Andreev bound states and the dispersion of the Andreev bound-state energies with the phase bias across the weak link.
In fact Andreev bound states play a central role in all transport properties of Josephson weak links, but as will be discussed below and in Sec. \ref{sec_Point-Contact-Heat-Transport}, their roles in charge and heat transport are fundamentally different. 
 
For a perfectly reflecting boundary with a small aperature of radius, $a \ll\xi_{\mbox{\tiny $\Delta$}}$, shown in Fig. \ref{fig_Josephson-Pinhole}, the total current transported through the weak link is so low that the phase gradient in the superconducting leads is negligible. Thus, for a superconducting leads specified by phases $\vartheta_L$ and $\vartheta_R$ asymptotically far from the aperature, the change in phase, $\vartheta$, 
occurs over a short distance $a\ll\xi_{\text{$\Delta$}}$ at the aperature.
For phase bias $\vartheta=\pi$ the point contact separates degenerate superconducting ground states with a pure sign change for each trajectory passing through the aperature.  
The resulting zero-energy bound states correspond to equal amplitude superpositions of normal-state electrons and and holes,
\be
\ket{\Psi_{\vp s}(\varepsilon=0)} = \sqrt{\frac{|\Delta|}{2\hbar\vv_{\vp}\cdot\hat\vn}}\,
\begin{pmatrix} i\,\chi_s \\ (i\sigma_y)\chi_s \end{pmatrix}
\,
e^{-|z|/\xi_{\text{$\Delta$}}}
\,.
\ee
If we tune the phase bias away from $\pi$ the bound state energy disperses with phase as shown in Fig. \ref{fig_Josephson-Current-Phase-Relation}. In addition there are positive and negative energy branches. The negative energy states are filled at zero temperature and responsible for the ground-state supercurrent at non-zero phase bias. At finite temperature thermal excitation of the Andreev bound sates leads to a reduction of the critical current.

In order to better understand the role of the bound state spectrum in charge and heat transport, as well as calculate these currents, we need to calculate the spectral functions for quasiparticles and Cooper pairs in the presence of boundaries, inhomogeneous pair potentials, as well the distribution functions for electron-like and hole-like excitations under non-equilibrium conditions. This requires dynamical equations for the propagators for quasiparticles and Cooper pairs, as well as boundary conditions connecting the scattering states involved in transport through the Josepshson weak link. 

\section{From Andreev to Eilenberger}\label{sec_Andreev-Eilenberger}

Andreev's equation is a transport-type equation for the evolution the Bogoliubov spinors 
along classical trajectories,
\be\label{eq-Andreev_Equation2}
\widehat{\cH}_{\text{A}}\,\ket{\Psi_{\vp}}
+
i\hbar\,\vv_{\vp}\cdot\grad\,\ket{\Psi_{\vp}}
= 0
\,,
\ee
with the operator $\cH_{\text{A}}$ defined by $\widehat{\cH}_{\text{A}}=\varepsilon\whtauz - \whDelta(\vr,\vp)$, where $\whDelta(\vr,\vp)$ is the Nambu matrix order parameter. For spin-singlet, s-wave pairing  
\be
\whDelta(\vr,\vp) = i\sigma_y\left(\whtaux\Delta_{1}(\vr) - \whtauy\Delta_{2}(\vr)\right)
\,,
\ee
where $\Delta_{1} = \Re\Delta=|\Delta|\cos{\vartheta}$ and 
$\Delta_{2} = \Im\Delta=|\Delta|\sin{\vartheta}$.

Andreev's equation can also be expressed in terms of a row spinor, 
\be
\bra{\tilde{\Psi}_{\vp}}\widehat{\cH}_{\text{A}} -
i\hbar\vv_{\vp}\cdot\grad\,\bra{\tilde{\Psi}_{\vp}} = 0
\,.
\ee
The row spinor $\bra{\tilde{\Psi}_{\vp}}$ is not simply the adjoint of $\ket{\Psi_{\vp}}$ since $\widehat{\cH}_{\text{A}}$ is not Hermitian, but is easily constructed, and the physical solutions are normalized to $\braket{\tilde{\Psi}_{\vp}}{\Psi_{\vp}}=1$.
For $|\varepsilon|>\Delta$, the two branches are propagating solutions, the particle-like
solution is $\ket{\Psi_{\vp}^{(+)}}$, with group velocity $\vv_{\vp} || \vp$, and the hole-like
solution is $\ket{\Psi_{\vp}^{(-)}}$, with reversed group velocity, $\vv_{\vp} || -\vp$.
For sub-gap energies the solutions are exploding and decaying functions along the trajectory, and thus relevant only in the vicinity of boundaries, domain walls, interfaces and weak links. 

The product of the particle- and hole amplitudes, 
\be
\cF_{\alpha\beta}(\vr,\vp;\varepsilon) 
= {u}_{\vp\alpha}(\vr;\varepsilon){v}_{\vp\beta}(\vr;\varepsilon)
\,,
\ee
is the \emph{Cooper pair propagator}, which determines the spectral composition of the pair potential. The latter satisfies the BCS mean-field self-consistency condition, which for spin-singlet pairing is 
\ber
\Delta_(\vr,\vp) 
=
\langle
g(\vp,\vp')
\fint
d\varepsilon
\tanh\left(\frac{\varepsilon}{2T}\right)
\cP(\vr,\vp';\varepsilon)
\rangle_{\vp'}
\,,
\eer
where the integration over the spectral function of correlated pairs, $\cP(\vr,\vp;\varepsilon)$, is cutoff at $\Omega_c\ll E_F$, the bandwidth of attraction for the pairing interaction, $g(\vp,\vp')$. The latter is integrated over the Fermi surface, $\langle\ldots\rangle_{\vp'}\equiv\int\,d\Omega_{\vp'}/4\pi(\ldots)$.

The pair propagator is one component of the Nambu matrix propagator,
\be
\whmfG(\vr,\vp;\varepsilon) = 
\sum_{\mu,\nu}\,g_{\mu\nu}\,\ket{\Psi_{\vp}^{(\mu)}}\bra{\tilde{\Psi}_{\vp}^{(\nu)}}
\,,
\ee
which satisfies Eilenberger's transport equation \cite{eil68},
\be\label{eq_Eilenberger-Equation}
\left[\widehat{\cH}_{\text{A}}\,,\,\whmfG(\vr,\vp;\varepsilon)\right] 
+ 
i\hbar\vv_{\vp}\cdot\grad\,\whmfG(\vr,\vp;\varepsilon) = 0
\,.
\ee
Physical solutions to Eq. \ref{eq_Eilenberger-Equation} must also satisfy Eilenberger's normalization condition \cite{eil68},
\be\label{eq_Normalization-condition}
\left(\whmfG(\vr,\vp;\varepsilon)\right)^2 = -\pi^2\widehat{1}
\,.
\ee

Eilenberger's formulation provides the spectral functions for both the quasiparticle and Cooper pair excitations from components of the quasiclassical propagator. For spin-singlet pairing in the absence of an external fields and magnetic interfaces the off-diagonal components of the propagator describe pure spin-singlet pairing correlations. 
As a result the Nambu propagator can be expressed in the form,
\be\label{eq-QCPropogator_Singlet}
\whGR = \cG^{\text{R}}\,\whtauz + i\sigma_y
	\left(\cF^{\text{R}}_{1}\whtaux - \cF^{\text{R}}_{2}\whtauy\right)
\,.
\ee
The superscript refers to the causal (retarded in time) propagator, obtained from Eq. \ref{eq_Eilenberger-Equation} with the shift, $\varepsilon\rightarrow\varepsilon+i0^{+}$.
The diagonal propagator in Nambu space, $\cG^{\text{R}}\whtauz$, determines the spectral function, or local density of states, for the fermionic excitations with momentum $\vp=p_f\hat{p}$, while the off-diagonal component, $\cF^{\text{R}} = \cF^{\text{R}}_{1}+i\cF^{\text{R}}_{2}$, determines the spectral function for the pairing correlations,
\be
\cN(\vr,\vp;\varepsilon) = -\frac{1}{\pi}\Im\,\cG^{\text{R}}(\vr,\vp;\varepsilon)
\,,\quad
\cP(\vr,\vp;\varepsilon) = -\frac{1}{\pi}\Im\,\cF^{\text{R}}(\vr,\vp;\varepsilon)
\,.
\ee

For spin-singlet pairing in the absence of magnetic fields and boundaries, we can transform the four-component Nambu matrices and spinors into two-component spinors in particle-hole space with a unitary transformation that removes the factors of $i\sigma_y$ corresponding to $\pi$ rotations of the hole spinors relative to the particle spinors, i.e. 
$\whDelta
=
(i\sigma_y)\left(\Re\Delta\whtaux - \Im\Delta\whtauy\right)
\rightarrow 
\widehat{U}_{\pi}^{\dag}\whDelta\widehat{U}_{\pi}
=
\left(\Re\Delta\whtaux + \Im\Delta\whtauy\right)
$, and 
$\whGR\rightarrow \widehat{U}_{\pi}^{\dag}\whGR\widehat{U}_{\pi}$ with 
$\widehat{U}_{\pi} = \mbox{diag} \left(\hat{1}\,,\,(-i\sigma_y)\right)$.

For homogeneous equilibrium the solution to Eilenberger's equation that also satisfies the normalization condition, and reproduces the normal-state density of states in the limit $\Delta=0$ is
\be
\whGR_{\text{$0$}}(\vp;\varepsilon) = -\pi\frac{\varepsilon\whtauz - \whDelta(\vp)}{\sqrt{(\varepsilon+i0^+)^2 - |\Delta(\vp)|^2}} 
\,,
\ee  
which describes the bulk spectral fuctions for quasiparticles and Cooper pairs,
\be
\cN(\vr,\vp;\varepsilon) = \frac{|\varepsilon|}{\sqrt{\varepsilon^2-|\Delta|^2}}\,\Theta(\varepsilon^2-|\Delta|^2)
\,,\quad
\cP(\vr,\vp;\varepsilon) = \frac{-\Delta}{\sqrt{\varepsilon^2-|\Delta|^2}}\,\Theta(\varepsilon^2-|\Delta|^2)
\,.
\ee

The equilibrium charge current in the quasiclassical theory can be calculated by weighting the charge currents of electrons and holes by their corresponding spectral weights and equlibrium occupations, 
\be
\vJ_e=2N_f\int\dangle{p}\int\ns d\varepsilon\,
\Big\{\ns
+e\vv^{(+)}_{\vp}\,\cN^{(e)}(\vr,\vp;\varepsilon)
\,f^{(e)}(\varepsilon)
-e\vv^{(-)}_{\vp}\,\cN^{(h)}(\vr,\vp;\varepsilon)
\,f^{(h)}(\varepsilon)
\Big\}
\,,
\label{eq_charge-current}
\ee
where $f^{(e)} = f(\varepsilon)$ is the equilibrium Fermi distribution for particle excitations and $f^{(h)}(\varepsilon) = 1-f(\varepsilon)$ is the equilibrium occupation for holes.
We can express the current in terms of the quasiclassical propagator by recognizing that the spectral weights for particle-like (hole-like) excitations for trajectory $\vp$ are given by $\cN^{(e)}=\cN^{(h)}=\cN(\vr,\vp;\varepsilon)=-\frac{1}{4\pi}\Im\Tr{\whtauz\whGR(\vr,\vp;\varepsilon)}$. Thus, the charge current reduces to 
\be
\vJ_e= 2N_f\int\dangle{\hat\vp}\int d\varepsilon\,e\vv_{\vp}\,
           \cN(\vr,\vp;\varepsilon)\,f(\varepsilon)
\,.
\label{eq_charge-current-GR-GA}
\ee
For equilibrium properties it is convenient to express the current in terms of the Matsubara propagator, which is related to the retarded and advanced propagators by analytic continuation, i.e. $\whGR(\varepsilon\rightarrow i\varepsilon_n) = \whGM(\varepsilon_n)$, where $i\varepsilon_n=i(2n+1)\pi T$ are the poles of the Fermi distribution. Thus, Eq. \ref{eq_charge-current-GR-GA} can be transformed to 
\be
\vJ_e = 2N_f\int\dangle{\hat\vp}\,e\vv_{\vp}\,T\sum_{n}\,
	  \nicefrac{1}{4}\Tr{\whtauz\whGM(\vr,\vp;\varepsilon_n)}
\,.
\label{eq_charge-current-GM}
\ee

\subsection{Sharvin Contact - Propagators and Spectral Functions}\label{sec_Point-Contact-Propagators}

Consider a Sharvin contact that couples two conventional, spin-singlet superconductors with phase bias, $\vartheta=\vartheta_R-\vartheta_L$.
For a trajectory $\vp$ passing through the aperature, as showin Fig. \ref{fig_Josephson-Pinhole}, the pair potentials of the two superconducting leads are given by, 
$\Delta(z < 0) \equiv \Delta^{(-)} = |\Delta|\,e^{-i\vartheta/2}$ and 
$\Delta(z > 0) \equiv \Delta^{(+)} = |\Delta|\,e^{+i\vartheta/2}$. In the limit, 
$a\ll\xi_0$, the phase change occurs at the point contact. 
The propagators far from the point contact are defined by local equilibrium propagators,
\be
\whGM_{\text{$0,\pm$}} = 
-\pi
\frac{i\varepsilon_n\whtauz -
	\left(\Delta^{(\pm)}_{1}
	\whtaux 
	-
	\Delta^{(\pm)}_{2}
	\whtauy
	\right)}
     {\sqrt{(\varepsilon_n^2 + |\Delta|^2}}
\,,
\label{eq_equilibrium-bulk-propagator}
\ee
where $\Delta^{(\pm)}_{1}=|\Delta|\cos(\vartheta/2)$ and $\Delta^{(\pm)}_{2}=\pm|\Delta|\sin(\vartheta/2)$.
The propagator obeys the Eilenberger equation, which is a first-order differential equation along the classical trajectory, and thus a continuous function of $x = \hat\vp\cdot\vr$. The inhomogeneity of the pair potential generates local solutions of Eq. \ref{eq_Eilenberger-Equation} that are confined within a few coherence lengths of the point contact, and encode the spectral information of the sub-gap Andreev states. 

It is convenient to transform the matrix transport equation to a linear differential equation acting on vectors defined in a 3-dimensional vector space \cite{sau11},
\be\label{eq_Eilenberger-Vector-Space}
\frac{1}{2}\vv_\vp \cdot \nabla \ket{\mfG} = \widehat{M} \ket{\mfG}\,
\ee
with 
\be
\ket{\mfG} \equiv 
\begin{pmatrix}
\cF_1	\\ \cF_2	\\ \cG	\\
\end{pmatrix}\, \quad 
\widehat{M} = 
\begin{pmatrix}
	0		&	-i\varepsilon_n	&	-\Delta_2	\\
	+i\varepsilon_n	&	0		&	 \Delta_1	\\
	-\Delta_2	& 	\Delta_1	&	0	
\end{pmatrix}
\,.
\ee
For either the left or right S region, we can express the physical solution $\ket{\mfG}$ in terms of the eigenvectors of the Hermitian matrix $\widehat{M}$, i.e. orthonormal solutions 
of $\widehat{M} \ket{\lambda} = \lambda\ket{\lambda}$.
There are three eigenvalues, $\left\{\lambda_{\mu} | \mu\in\{0,\pm 1\}\right\}$:
$\lambda_0 = 0$ and $\lambda_{\pm} = \pm\lambda$, with 
$\lambda = \sqrt{\varepsilon_n^2 + |\Delta|^2}$,
The eigenvector with eigenvalue $\mu =0$ is
\be
\ket{0} = \frac{1}{\lambda}
\begin{pmatrix}
	\Delta_1	\\
	\Delta_2	\\
	-i\varepsilon_n
\end{pmatrix}
\,,
\ee
and corresponds to the bulk equilibrium propagator in Eq. \ref{eq_equilibrium-bulk-propagator}.
The eigenvectors for eigenvalues $\lambda_{\pm} = \pm \lambda$ are 
\be
\ket{\pm} = \frac{1}{\sqrt{2}\lambda \lambda_1} 
\begin{pmatrix}
	\pm i\varepsilon_n\lambda - \Delta_1 \Delta_2	\\
	\lambda_1^2					\\
	i\varepsilon_n \Delta_2 \mp \lambda \Delta_1
\end{pmatrix}
\,,
\ee
where $\lambda_1 \equiv \sqrt{\varepsilon_n^2 + \Delta_1^2}$. The set of eigenvectors are orthonormal, $\braket{\mu}{\nu} = \delta_{\mu\nu}$.
The general solution in half-space $x < 0$ ($x > 0$) can be expressed in terms of eigenvectors of the corresponding half-space operator, $\widehat{M}^{\text{$\pm$}}$,
\be
\ket{\mfG^{(\text{$\pm$})}(x)} 
= 
\sum_{\mu}\,C_{\mu}^{\text{($\pm$)}}(x)\,\ket{\mu}^{\text{$\pm$}}
\,.
\ee
From Eq. \ref{eq_Eilenberger-Vector-Space} we project out the differential equations for the amplitudes,
\be
\nicefrac{1}{2}v_f\pder{C^{(\text{$\pm$})}_{\mu}(x)}{x}=\lambda_{\mu}\, C^{\text{$(\pm)$}}_{\mu}(x)
\,.
\ee
Thus, there is a constant solution for $\mu=0$, and exponential solutions for $\mu=\pm$,
\ber
C^{\text{$(\pm)$}}_{0}(x) = C^{\text{$(\pm)$}}_{0}(0)
\,,
\quad
C^{\text{$(\pm)$}}_{+}(x) = C^{\text{$(\pm)$}}_{+}(0)\,e^{+2\lambda x/v_f}
\,,
\quad
C^{\text{$(\pm)$}}_{-}(x) = C^{\text{$(\pm)$}}_{-}(0)\,e^{-2\lambda x/v_f}
\,.
\eer

\begin{figure}[t]
\centering
\begin{tabular}{ll}
\raisebox{0.5\totalheight}{\hspace*{-5mm}
\includegraphics[width=0.5\textwidth]{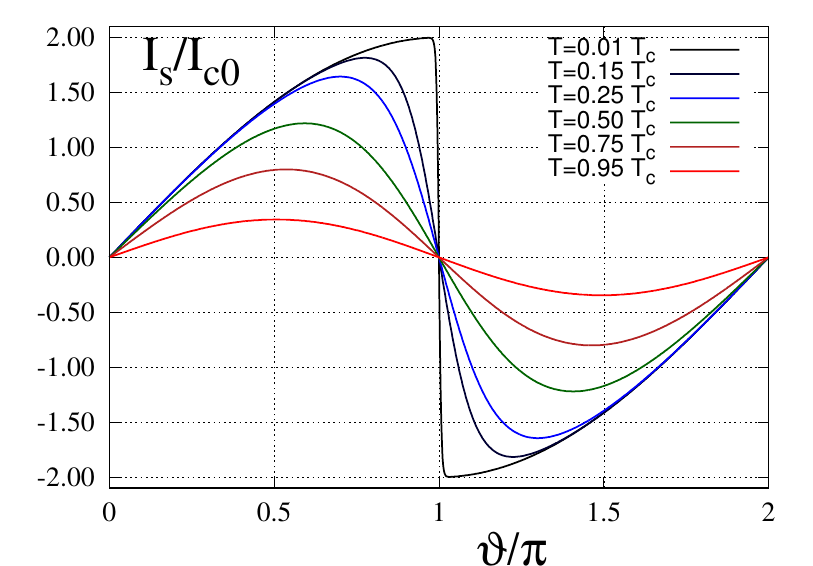}
}
&
\raisebox{0.5\totalheight}{\hspace*{-5mm}
\includegraphics[width=0.5\textwidth]{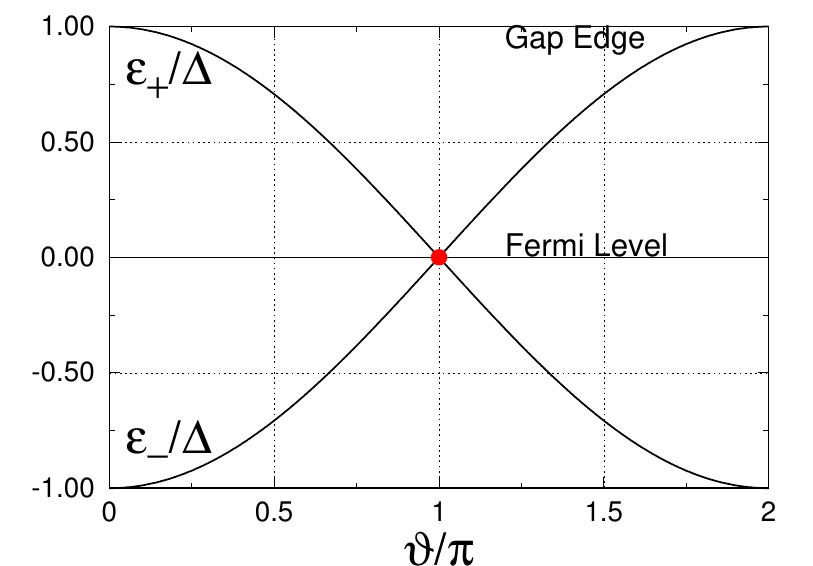}
}
\end{tabular}
\caption{
Left: Josephson current as a function of phase bias and temperature for a Sharvin point contact. The current is normalized in units of the Ambegaokar-Baratoff critical current at $T=0$, $I_{c_0}=\frac{\pi\Delta(0)}{2e\,R_{\text{N}}}$.
Right: Branches of Andreev bound states of a Sharvin point-contact Josephson weak link dispersing as a function of the phase bias, $\vartheta$.
}
\label{fig_Josephson-Current-Phase-Relation}
\end{figure}

The half-space solutions must asymptotically approach the bulk solution, and thus take the form,
\ber
\ket{\mfG^{(\text{$-$})}(x)} 
&=& 
\pi\,
\ket{0}^{\text{$-$}} 
+ 
C^{\text{$(-)$}}_+\,e^{+2\lambda x/v_f}\ket{+}^{\text{$-$}}
\,,\quad x < 0
\\
\ket{\mfG^{(\text{$+$})}(x)} 
&=& 
\pi\,
\ket{0}^{\text{$+$}} 
+ 
C^{\text{$(+)$}}_-\,e^{-2\lambda x/v_f}\ket{-}^{\text{$+$}}
\,,\quad x > 0
\,.
\eer
Continuity of the propagator at $x=0$ fixes the amplitudes, 
$C^{\text{$(-)$}}_{+}=-\sqrt{2}\pi\nicefrac{\Delta^{\text{$(-)$}}_2}{\lambda_{1}}=-C^{\text{$(+)$}}_{-}$. Thus, 
the quasiparticle component of the propagator evaluated at the point contact beomes,
\be
\cG(\vp,z=0^-;\varepsilon_n)
=-\pi\frac{i\varepsilon_n\lambda-\nicefrac{1}{2}|\Delta|^2\,\sin\vartheta}
                       {\varepsilon_n^2 + |\Delta|^2\,\cos^2(\vartheta/2)}
\ee
The total supercurrent at the point contact is obtained from Eq. \ref{eq_charge-current-GM} and the 
cross-section of the Sharvin contact,
\be
I_s = (\pi a^2)\,2N_f\int_{\hat\vp\cdot\hat\vn>0}\ms\dangle{\vp}\,2\hat\vn\cdot\vv_p\,
      \frac{\pi}{2}\,T\sum_{n}\,\frac{|\Delta|^2\sin\vartheta}{\varepsilon_n^2 + |\Delta|^2\cos^2(\vartheta/2)}
\,.
\label{eq_Current-KO-1}
\ee
The normal-state Sharvin conductance, $G_{\text{N}} = \nicefrac{\pi a^2}{2} e^2\,N_f v_f$, valid in the ballistic limit $\ell \gg a$, is determined by the area of the aperature, the density of states and Fermi velocity for quasiparticle charge transport. Thus, in terms of the contact resistance, $R_{\text{N}} = 1/G_{\text{N}}$,
\be
I_s 
= 
\frac{\pi}{e R_{\text{N}}}
T\sum_{n}\,\frac{|\Delta|^2\sin\vartheta}{\varepsilon_n^2+|\Delta|^2\cos^2(\vartheta/2)}
=
\frac{\pi|\Delta|}{e\,R_{\text{N}}}\,\sin(\vartheta/2)
   \tanh\left(\frac{|\Delta|\cos(\vartheta/2)}{2T}\right)
\,,
\label{eq_Current-KO-2}
\ee
where the last form is obtained from the Matsubara series representation for the hyperbolic tangent function.
This is the result for the current-phase relation originally obtained by K-O \cite{kul77}. It reduces to the current-phase relation for a Josephson tunnel junction, $I_s=I_c\sin\vartheta$, in the limit $T\rightarrow T_c$, where the critical current,
\be
I_c = \frac{\pi|\Delta|^2}{4e\,R_{\text{N}}T_c}\,\propto (1-T/T_c)\,,\quad |T - T_c| \ll T_c
\,,
\ee
is the result obtained by Ambegaokar and Baratoff \cite{amb63}, except that the tunnelling conductance is replaced by the Sharvin conductance,
However, at low temperatures the current-phase becomes increasingly asymmetric near $\vartheta = \pi$ as shown in Fig. \ref{fig_Josephson-Current-Phase-Relation}, and discontinuous at $T=0$, 
\be
I_s(T=0)=\frac{\pi\Delta(0)}{e\,R_{\text{N}}}\,\sin(\vartheta/2)\,\sgn(\cos(\vartheta/2))
\,.
\ee
In addition to the divergence of the slope $dI_s/d\vartheta|_{\vartheta=\pi}$ as $T\rightarrow 0$, the K-O result predicts a critical current that is twice that of the Ambegaokar-Baratoff result.   

The anomaly in the temperature dependence of the current-phase relation of the Josephson current obtained by Kulik-Omelyanchouck has its origin in the Andreev bound-state spectrum of the phase-biased Sharvin weak link. This can be made clear by expressing the current defined by Eq. \ref{eq_Current-KO-1} as an integration over the spectral current density, weighted by thermal occupation of the current-carrying excitation spectrum. 
Starting from the first equality of Eq. \ref{eq_Current-KO-2}, the current-phase relation is determined by the sum,
\be
J = T\sum_n\frac{\Delta_1\Delta_2}{\varepsilon_n^2 + \Delta_1^2} \equiv T\sum_n\,G(i\varepsilon_n)
\,,
\ee
where $\Delta_1 = |\Delta|\cos(\vartheta/2)$ and $\Delta_2 = |\Delta|\sin(\vartheta/2)$, and $G(z) = -\Delta_1\Delta_2/(\Delta_1^2 - z^2)$ is the analytic extension of the summand to the complex plane. 
We can transform the current-phase relation to an integral over the spectral density for the current by 
expressing the Matsubara sum as a contour integral around the poles of the Fermi function, use analyticity 
to deform the contour to integration infinitesimally above and below the real axis, 
\be
J = \int_{-\infty}^{+\infty} d\varepsilon\,\left(f(\varepsilon)-\frac{1}{2}\right)\,
\left(\frac{G^R(\varepsilon) - G^A(\varepsilon)}{-2\pi i}\right)
\,,
\ee 
where $G^R(\varepsilon) = G(\varepsilon + i0^+)$ ($G^A(\varepsilon) = G(\varepsilon - i0^+)$) is the retarded (advanced) current response to the phase bias, and the difference of these two repsonse functions is the spectral function for the Josephson current  
\ber
\Gamma(\varepsilon) =  \frac{G^R(\varepsilon) - G^A(\varepsilon)}{-2\pi i} 
=
\frac{1}{2}\Delta_2\,
\left[\delta(\varepsilon-\varepsilon_{-}(\vartheta))-
      \delta(\varepsilon-\varepsilon_{+}(\vartheta))
\right]
\,.
\eer
The spectral function is defined by delta functions at energies corresponding to the positive and negative energy Andreev levels,
\be
\varepsilon_{\pm}(\vartheta) = \pm|\Delta|\cos(\vartheta/2)
\,,\quad -\pi\le\vartheta\le+\pi
\,.
\ee
The dispersion of the Andreev levels with phase is shown in the right panel of Fig. \ref{fig_Josephson-Current-Phase-Relation}
The phase gradient of these energy levels 
\be
I_{\pm}(\vartheta) = \der{\varepsilon_{\pm}(\vartheta)}{\vartheta} 
		   = \mp\frac{1}{2}|\Delta|\sin(\vartheta/2)
\ee
determines the spectral weight of the Andreev bound states to the Josephson current.
These gradients determine the magnitude of the current transported by the Andreev levels in the Sharvin aperature. In particular, the current spectral function is 
\ber
\Gamma(\varepsilon) =  
\left[I_{-}(\vartheta)\delta(\varepsilon-\varepsilon_{-}(\vartheta))
      +
      I_{+}(\vartheta)\delta(\varepsilon-\varepsilon_{+}(\vartheta))
\right]
\,.
\eer
With this result for the spectral current density the Josephson current becomes,
\be
I_s 
=
\frac{\pi}{e\,R_{\text{N}}}
\sum_{\nu = \pm}\,f(\varepsilon_{\nu})\,\der{\varepsilon_{\nu}}{\vartheta}
\,.
\label{eq_point-contact-CPR}
\ee
Evaluating this result reduces to the K-O result for the current-phase relation in Eq. \ref{eq_Current-KO-2}, and demonstrates explicitly that the Josephson current of a Sharvin point-contact is transported by the occupied spectrum of Andreev bound states (see also Ref. \cite{bag92}). In particular, the Josephson current at $T=0$ is determined by the occupied spectrum of negative energy Andreev levels, and on this basis the discontinuity in the current-phase relation at $\vartheta = \pi$ is explicit from the discontinuity in the slope of the negative energy branch shown in the right panel of Fig. \ref{fig_Josephson-Current-Phase-Relation}.

\subsection{Partially Transparent Point Contacts}\label{sec_Point-Contact-Transparency} 

\begin{figure}[t]
\centering
\begin{tabular}{lll}
\raisebox{0.10\totalheight}{\hspace*{-5mm}
\includegraphics[width=0.45\textwidth]{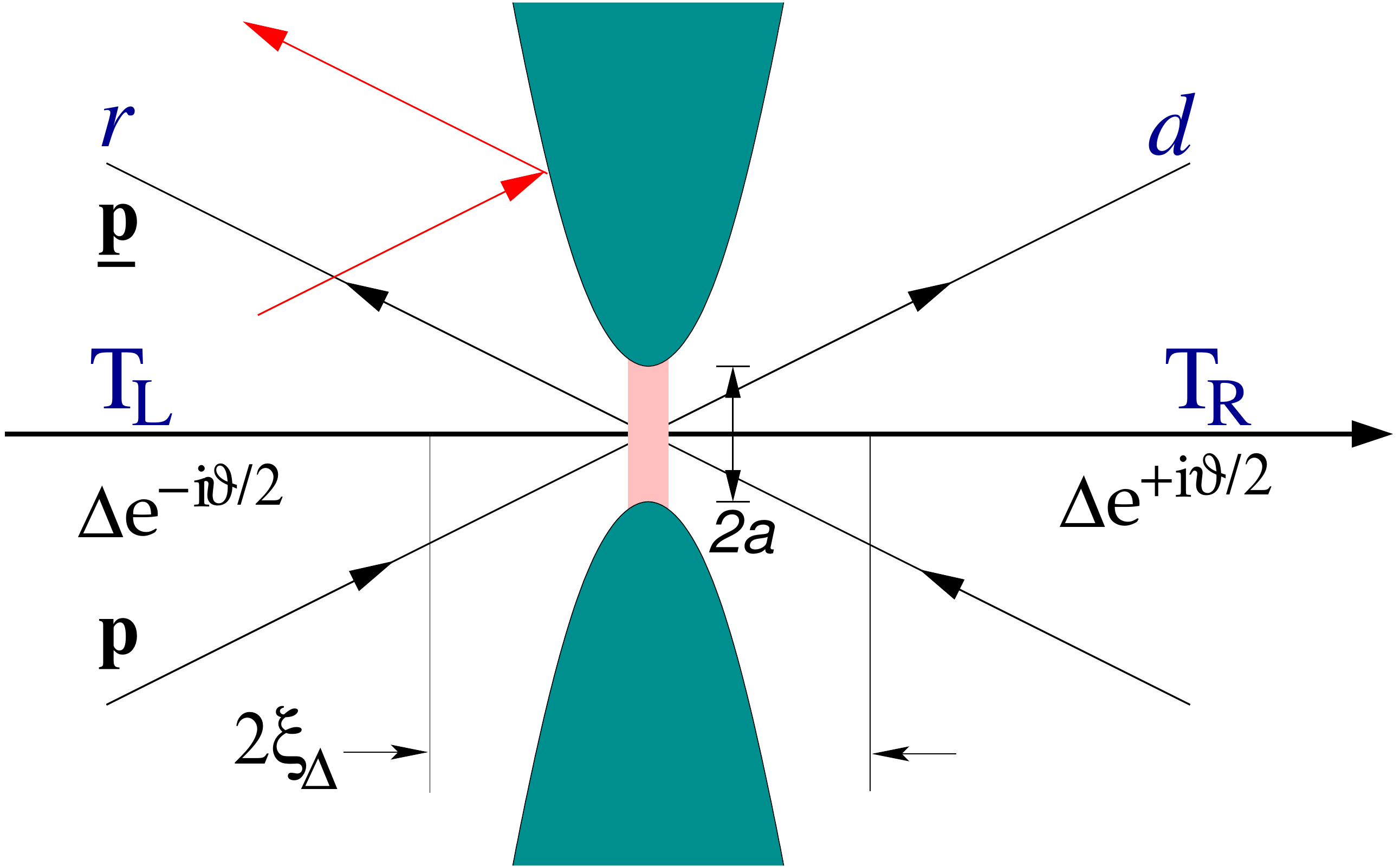}
}
&
\qquad
&
\includegraphics[width=0.45\textwidth]{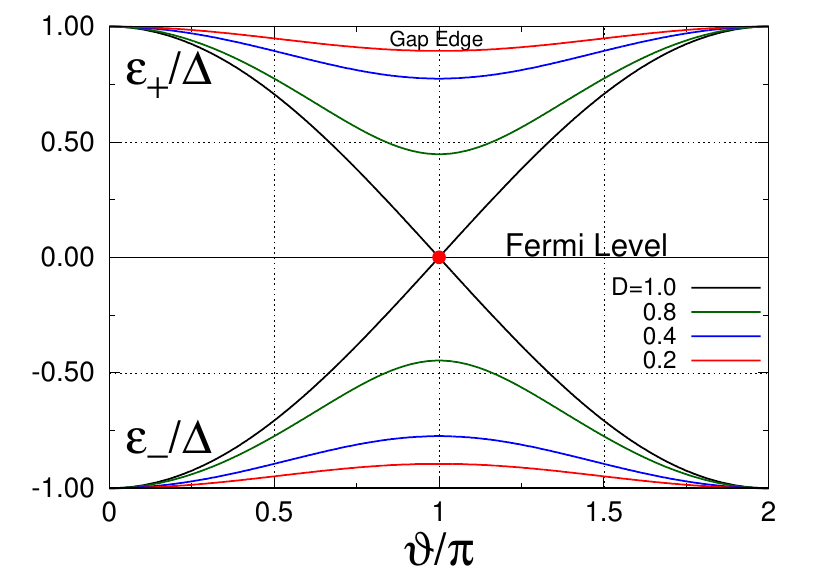}
\end{tabular}
\caption{
Left: Josephson point contact with radius $a\ll\xi_{\text{$\Delta$}}$. Current is transported only via trajectories passing through the aperature with transmission probability $D$. Right: Branches of Andreev bound states of an S-c-S weak link dispersing as a function of the phase bias, $\vartheta$.
}
\label{fig_Josephson-Point_Contact_S-matrix}
\end{figure}

The point-contact Josephson weak link discussed above is based on perfect transmission for classical trajectories that pass through the aperature of the S-S contact. 
The generalization of the Sharvin contact to a partially transparent contact requires boundary conditions for the propagators corresponding to scattering states related by the normal-state scattering matrix (S-matrix) as shown in Fig. \ref{fig_Josephson-Point_Contact_S-matrix}, 
\be
S = \begin{pmatrix} r & d \\ d^* & r \end{pmatrix}
\,.
\ee
where $r$ ($d$) is the reflection (transmission) amplitude of normal-state quasiparticles for trajectories passing through the aperature from the left to the right lead. Unitarity of the S-matrix requires $|r|^2 + |d|^2 = 1$, where $R = |r|^2$ ($D=|d|^2$) is the probability for reflection (transmission) by the point contact.
The reflection and transmission amplitudes depend on the incident, reflected and transmitted trajectories, $\vp$ and $\underline\vp$, shown in the left panel of Fig. \ref{fig_Josephson-Point_Contact_S-matrix}. For a non-magnetic interface $r$ and $d$ are independent of the quasiparticle spin. In the following we assume the S-matrix amplitudes are independent of the 
momenta, except the S-matrix couples only states that conserve momentum parallel to the 
interface.

Interface scattering couples states defined on different trajectories, and when combined with Andreev scattering, modifies the spectrum of Andreev bound states at the point contact. The theory of Andreev bound-state formation at partially transparent interfaces, with multiple scattering by the boundary and pair potential, is formulated in terms of boundary conditions for the particle-hole coherence amplitudes that can be used to construct the quasiclassical propagators \cite{kur87,esc00,zha08}. The formulation of these boundary conditions and generalizations to magnetically active interfaces, with applications to pair breaking in unconventional superconductors, superfluid \He\ and hybrid superconducting-magnetic materials, is discussed by M. Eschrig \cite{esc18}.

For a partially transparent, non-magnetic point contact that couples only the incident, specularly reflected and momentum conserving transmitted trajectory the Andreev bound energies are given by \cite{zha03},
\ber
\varepsilon_{\pm}(\vartheta,R)=\pm|\Delta|\,\sqrt{\cos^2(\vartheta/2)
                              +R\,\sin^2(\vartheta/2)}
\,.
\label{eq_point-contact-dispersion}
\eer
Thus, boundary reflection opens a low-energy gap in the Andreev bound-state spectrum at 
$\vartheta=\pi$, $\varepsilon_{+}-\varepsilon_{-}=2R\,|\Delta|$.
Thus, there is no topological protection of the zero mode for $\vartheta=\pi$ because there is a finite probability for back reflection with no change in phase.

\begin{figure}[t]
\centering
\includegraphics[width=0.7\textwidth]{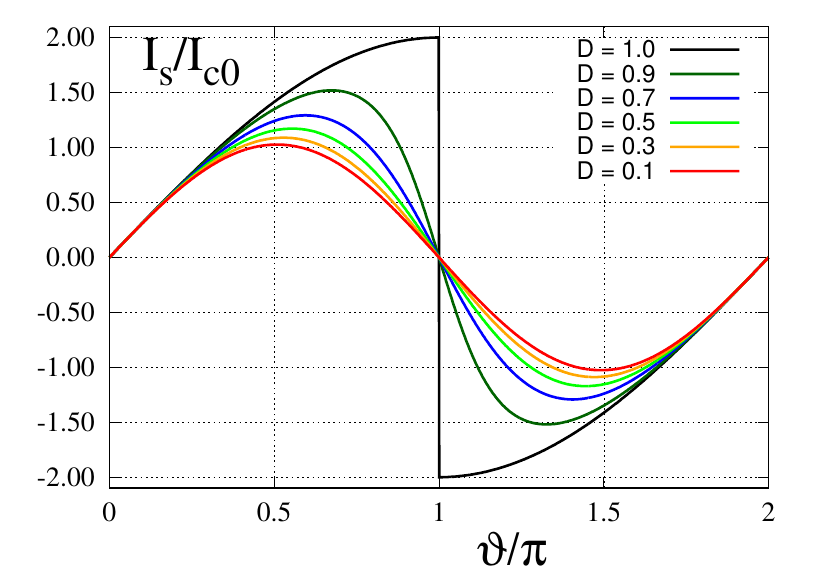}
\caption{
Josesphson current-phase relation for an S-c-S point-contact at $T=0$ as a function of the 
transmission probability, $0 < D \le 1$. The current is normalized by the Ambegaokar-Baratoff critical current at $T=0$, $I_c^{\text{AB}} = (\pi|\Delta|/2eR_{\text{N}}) \times D$.
}
\label{fig_point-contact-CPR}
\end{figure}

For a small area point contact with finite transmission the Josephson current is determined by Eq. \ref{eq_point-contact-CPR}, but with the the Andreev bound-state dispersion given by Eq. \ref{eq_point-contact-dispersion}. 
In the zero-temperature limit the current is carried entirely by the negative energy branch in Fig. \ref{fig_Josephson-Point_Contact_S-matrix},
\be
I_s(0,D) = 
I_{c}^{\text{AB}}(D) 
\frac{\sin\vartheta}{\sqrt{\cos^2(\vartheta/2) + R\,\sin^2(\vartheta/2)}}
         = 
	\left\{
	\begin{tabular}{ll}
	$I_{c}^{\text{AB}}(0)\,\sin\vartheta\,,$ 
	& 
	$D\rightarrow 0\,,$
	\\
	$2I_{c}^{\text{AB}}(1)\,\sin(\vartheta/2)\,\mbox{\it sgn}\left(\cos(\vartheta/2)\right)\,$ 
	& 
	$D\rightarrow 1\,,$
	\end{tabular}
	\right.
\label{eq_point-contact-current-T=0}
\ee
which reduces to the result obtained by Ambegaokar and Baratoff for a tunnel junction in the limit $D\ll 1$, with 
$I_{c}^{\text{AB}}(D) = \nicefrac{\pi|\Delta|}{2eR^{\text{AB}}_{\text{N}}}$, where $1/R^{\text{AB}}_{\text{N}}\equiv 1/R_{\text{N}}\times D$ is equivalent to the perturbation theory result for the tunnelling conductance between normal-metal leads in the limit $D\rightarrow 0$, and the Sharvin conductance in the limit $D\rightarrow 1$.
For transparency $D\rightarrow 1$ we recover the result of Kulik and Omelyanchouck with an 
enhanced critical current and the discontinuity in the slope associated with the Andreev bound-state dispersion near $\vartheta=\pi$.   
Figure \ref{fig_point-contact-CPR} shows the evolution of the Josephson current-phase relation with the the barrier transparency, $D$, at zero temperature.  
The effect of a gap in the bound state spectrum is to ``normalize'' the current-phase relation towards the Ambegaokar-Baratoff result for the Josephson current-phase relation based on second-order perturbation theory in the tunneling Hamiltonian.

The approach of the critical current to that obtained by Ambegaokar and Baratoff might suggest that perturbation theory in the transmission probability, $D$, is generally valid to a broader range of transport processes in superconducting point contacts.
This however is not the case. Even very low transmission, $D\ll 1$, leads to Andreev bound-state formation. The multiple scattering processes that lead to bound-state formation also affect the continuum spectrum near the gap edge. 
This process is inherently \emph{non-perturbative}, and when the back-action of bound-state formation on the continuum spectrum is relevant to transport the non-perturbative nature of the Andreev bound-state spectrum is revealed. There are many examples in which the back-action on the continuum spectrum resulting from Andreev bound-state formation leads to important new physics. In this volume Mizushima and Machida discuss the importance of the back-action by the continuum on the edge current in chiral superfluids and superconductors, and in particular the magnitude of the ground-state angular momentum of superfluid \Hea. The edge current is determined by both the negative-energy Andreev bound states which are chiral Fermions, and the current generated by the back-action on the continuum spectrum \cite{sto04,sau11,tsu12}.
In the context of phase-biased Josphson junctions and weak links, the non-perturbative nature of transport in superconducting junctions is particularly evident when considering heat transport through Josephson junctions. 

\section{Heat Transport in Josephson Junctions and Point Contacts}
\label{sec_Point-Contact-Heat-Transport}

Heat is carried by unbound quasiparticles making up the continuum spectrum above the gap of an S-c-S Josephson junction, and it is the imprint of the Andreev bound-state spectrum on the transmission probability of these quasiparticles through an S-c-S junction that reveals the non-perturbative nature of transport in Josephson junctions, and in this case the break-down of perturbation theory based on the tunneling Hamiltonian for heat transport, as well as other non-equilibrium processes such as thermo-electric transport \cite{gut97a}, and heat current noise \cite{gol13}. Here I focus on heat transport in phase-biased and temperature-biased Josephson junctions and point contacts. 

Heat transport through a Josephson junction was first studied by Maki and Griffin \cite{mak65}, and in more detail by Guttman et al. \cite{gut97,gut98}, following closely the perturbation theory approach based on the tunneling Hamiltonian method employed by Josephson \cite{jos62}, Ambegoakar and Baratoff \cite{amb63}. The failure of perturbation theory in the tunneling Hamiltonian (tH) is evident in all three papers.
In particular, the authors of Ref. \cite{gut98} give the following result for the heat current transported through a Josephson junction with temperature bias $\delta T = T_{R} - T_{L}$,\footnote{Note that I have used the notation of this article for the normal-state transmission probability and normal-state density of states factors, etc. What matters here is the integral over the joint density of states for the bulk quasiparticle excitations and coherence factors which is as given in Ref. \cite{gut98}.}
\begin{equation}
I_{\text{Q}}^{\text{tH}} = - \cA N_f v_f\,D\,\delta T
                          \int_{\Delta(T)}^{\infty}d\varepsilon\,
			  \frac{\varepsilon^2}{T}
                          \left(-\pder{f}{\varepsilon}\right)
                          \frac{\varepsilon^2-\Delta(T)^2\cos\vartheta}
                               {\sqrt{\varepsilon^2-\Delta(T+\delta T)^2}\sqrt{\varepsilon^2-\Delta(T)^2}}
                          \,.
\label{eq_heat-current-tH}
\end{equation}
For any phase bias, $\vartheta\ne 0,2\pi$, the tH result is divergent due to an essential singularity at $\varepsilon=\Delta$ in the linear response limit, i.e. to linear order in $\delta T$. The authors of Refs. \cite{gut98} and \cite{mak65} regulate the divergence in the formula for the heat current of the tH theory by introducting the temperature bias into the density of states of the higher temperature superconducting lead. This adhoc procedure leads to a non-analytic dependence of the heat current on the temperature bias, and thus to a failure of linear response theory for heat transport through a Josephson junction with arbitrarily small temperature bias $\delta T$. 
%
However, the resolution of the singularity in the tunneling conductance is not found in the breakdown of linear response theory, but in a failure of perturbation theory in the tunneling Hamiltonian. Indeed the result of the tH theory is missing the effects of Andreev bound-state formation, which is non-perturbative, on the continuum spectrum, and specfically on the transmission probability of the Bogoliubov quasiparticles that transport thermal energy.

A non-perturbative formulation of the theory of heat transport in Josephson junctions based on an S-matrix theory for multiple barrier and Andreev scattering was presented in Refs. \cite{zha03,zha03a}, and extended to charge- spin-transport transport under non-equilibrium conditions in Refs. \cite{esc03,zha07,zha08}, and described by Eschrig \cite{esc18} and Holmqvist et al. \cite{hol18}. 

The heat transported through an S-c-S Josephson junction or point-contact is carried by Bogoliubov particle and hole-like excitations in the superconducting leads. Thus, for weak non-equilibrium conditions the heat current is linear in the temperature bias, $J_{\text{Q}} = -\kappa\,\delta T$, with the thermal conductance of the S-c-S junction given by
\be
\kappa = \cA \int_{|\Delta|}^{\infty}d\varepsilon\,
             \cN_{\text{B}}(\varepsilon)\left[\varepsilon\,
	     v_{\text{B}}(\varepsilon)\right]\,
	     \cD(\varepsilon,\vartheta)\,
	     \pder{f}{T}
       = \cA\,N_f\,v_f \int_{|\Delta|}^{\infty}d\varepsilon\,
	     \cD(\varepsilon,\vartheta)\,
	     \left(\frac{\varepsilon^2}{T}\right)
	     \left(-\pder{f}{\varepsilon}\right)
\,,
\label{eq_thermal-conductance-ScS}
\ee
where $\partial f/\partial T = (-\varepsilon/T)\partial f/\partial \varepsilon$ is the thermal bias of the quasiparticle distributions of the superconducting leads. The second form of Eq. \ref{eq_thermal-conductance-ScS} follows from the density of states for the quasiparticles in the superconducting leads, $\cN_{\text{B}}(\varepsilon) = N_f\,\varepsilon/\sqrt{\varepsilon^2-|\Delta|^2}$, and the speed of these excitations, $v_{\text{B}}(\varepsilon) = v_f\, \sqrt{\varepsilon^2-|\Delta|^2}/\varepsilon$. 
The key element in Eq. \ref{eq_thermal-conductance-ScS} is the transmission probability,
$\cD(\varepsilon,\vartheta)$, for Bogoliubov quasiparticles with energy $\varepsilon$ in the presence of the phase bias, $\vartheta$. The normal-state transmission probability, $D$, is renormalized by multiple barrier and Andreev scattering from the phase-biased order parameter.

\begin{figure}[t]
\centering
\includegraphics[width=0.45\textwidth]{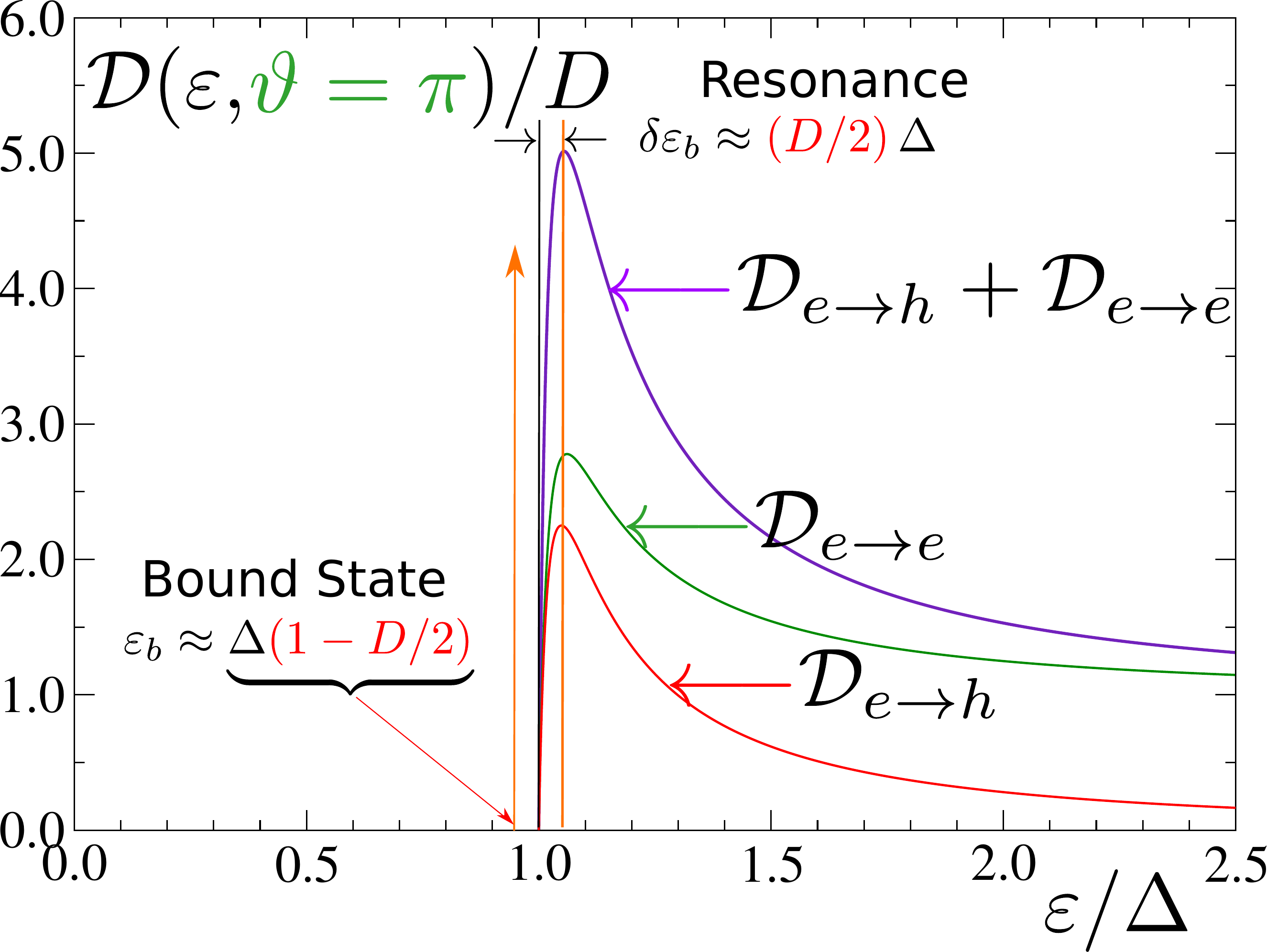}
\quad
\includegraphics[width=0.45\textwidth]{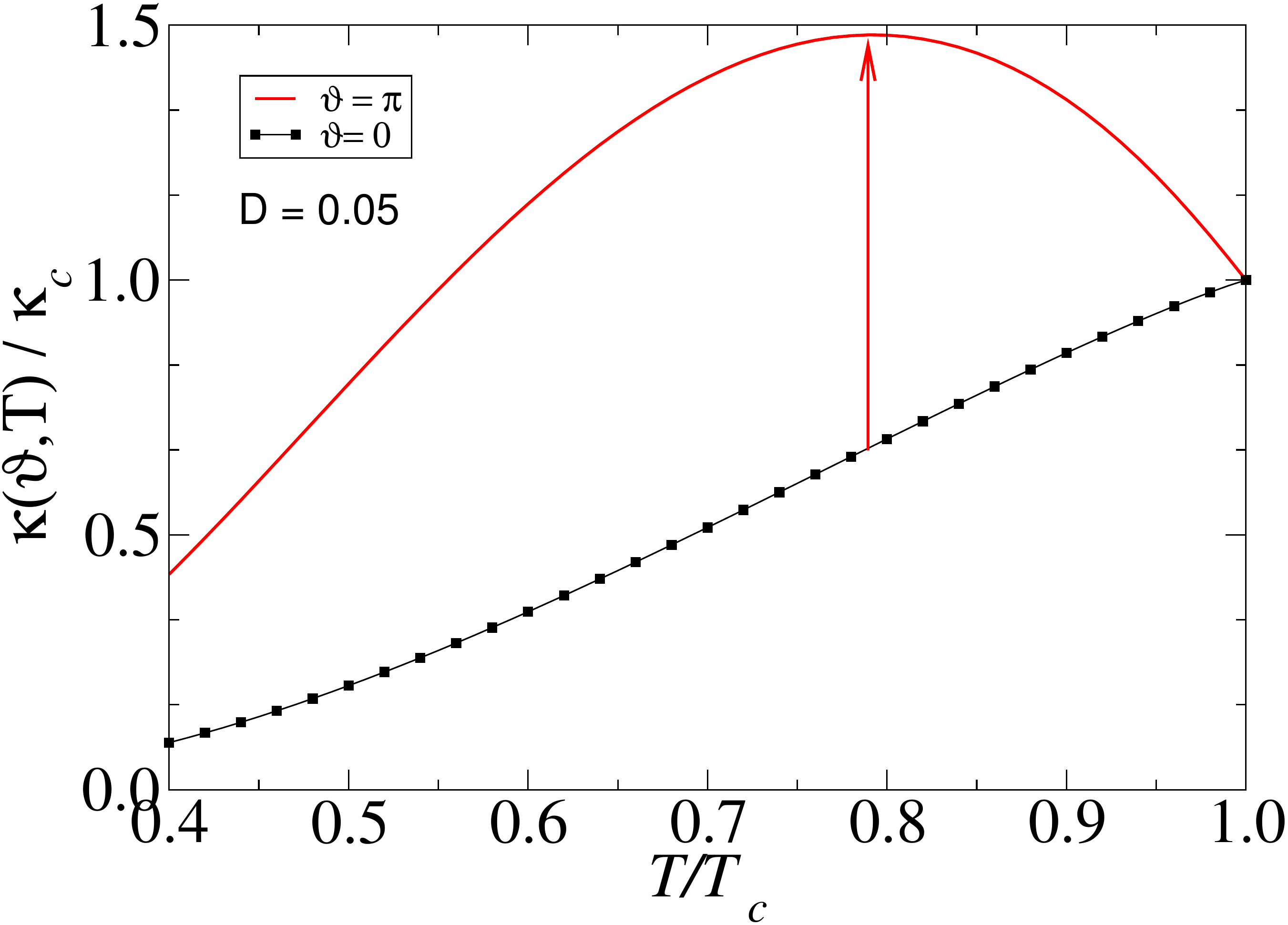}
\caption{
Left panel: 
Resonant quasiparticle transmission for an S-c-S Josephson junction tuned to $\vartheta=\pi$ and $D=0.2$. The Andreev bound state lies just below the continuum with $\varepsilon_{\text{b}} \simeq |\Delta| (1 - D/2)$, giving rise to a transmission resonance at an energy $\varepsilon_{\text{res}} \simeq |\Delta| (1 + D/2)$. 
Right panel: 
Thermal conductance of a phase-biased ScS Josephson weak link. For $\vartheta =0$ (black curve) the thermal conductance is suppressed below 
$T_c$. For phase bias $\vartheta=\pi$ heat conduction \emph{increases} sharply below $T_c$ reflecting resonant transmission of hot quasiparticles due to resonant transmission generated by a shallow Andreev bound state \cite{zha03}.
}
\label{fig_heat-conductance-Josephson-Point-Contact}
\end{figure}

\subsection{Phase-Sensitive Heat Transport - Andreev's Demon}

Multiple barrier and Andreev scattering leads to the renormalized transmission probability,
$\cD = \cD_{\text{$e\rightarrow e$}}+\cD_{\text{$e\rightarrow h$}}$, which is the sum of the probabilities for transmission with ($e\rightarrow h$ and $h\rightarrow e$) and without ($e\rightarrow e$ and $h\rightarrow h$) branch conversion \cite{zha03,zha03a},
\ber
\cD_{
	\text{$e\rightarrow e$}
	\atop
	\text{$h\rightarrow h$}
	} 
&=& 
D\,\,\,
\frac{(\varepsilon^2-|\Delta|^2)(\varepsilon^2-|\Delta|^2\cos^2(\vartheta/2))}
     {[\varepsilon^2-|\Delta|^2(1-D\sin^2(\vartheta/2))]^2}
     \,,\quad \varepsilon \ge |\Delta|
\,,
\label{eq_D-ee}
\\
\cD_{
	\text{$e\rightarrow h$}
	\atop
	\text{$h\rightarrow e$}
	}
&=& 
DR
\frac{(\varepsilon^2-|\Delta|^2)\,|\Delta|^2\sin^2(\vartheta/2)}
     {[\varepsilon^2-|\Delta|^2(1-D\sin^2(\vartheta/2))]^2}
     \,,\quad \varepsilon \ge |\Delta|
\,.
\label{eq_D-eh}
\eer
Both transmission amplitudes have poles at sub-gap energies corresponding to the Andreev bound-state energies given in Eq. \ref{eq_point-contact-dispersion}. 
Transmission with branch-conversion ($\cD_{\text{$e\rightarrow h$}}$) vanishes in the limit $\vartheta=0$, as does the sub-gap Andreev bound-state. The transmission probability is then given by the normal-state transparency. As a result Eq. \ref{eq_thermal-conductance-ScS} for the thermal conductance reduces to that for the S-S contact in Eq. \ref{eq-conductance-AR}, i.e. without Andreev reflection.
Transmission with branch-conversion also vanishes for a Sharvin contact without barrier reflection. Nevertheless, the transmission probability for Bogoliubov excitations is renormalized by the formation of the point-contact Andreev bound state by multiple Andreev scattering by the phase-change of the pair potential,
\be
\cD = \frac{(\varepsilon^2-|\Delta|^2)}
     {\varepsilon^2-|\Delta|^2\cos^2(\vartheta/2)}\,,\quad \varepsilon \ge |\Delta|
\,,
\ee
which gives the result for the heat current of a superconducting S-c-S contact first obtained by Kulik and Omelyanchouck \cite{kul92}, and which is 
analytic for $\vartheta\ne 0$ and $\delta T\rightarrow 0$. Both the phase dependence of the heat current, and the analyticity for small thermal bias,
originate from the appearance of the Andreev bound-state energy in the denominator of the transmission probability for Bogoliubov quasiparticle excitations.

Indeed for thermal excitations, i.e. energies $\varepsilon\ge|\Delta|$ above the gap, the Andreev bound-state pole (i) eliminates the unphysical divergence that is present in the perturbative result for the thermal conductance based on the tunneling Hamiltonian, and (ii) for low normal-state barrier transmission, $D\ll 1$, and phase bias tuned to $\vartheta=\pi$, the shallow Andreev bound-state leads to \emph{resonant} heat transport by quasiparticles.
The resonance in the transmission probability for $\vartheta = \pi$ is shown in Fig. \ref{fig_heat-conductance-Josephson-Point-Contact}.
The Andreev bound state lies just below the continuum edge, 
$\varepsilon_{\text{b}} \simeq |\Delta|(1 - D/2)$, 
and generates a transmission resonance peak just above the gap,
$\varepsilon_{\text{res}} \simeq |\Delta|(1 + D/2)$.


It is worth noting that in the limit $D\ll 1$, if we set $D=0$ everywhere except for the prefactor in Eqs. \ref{eq_D-ee} and \ref{eq_D-eh}, i.e. evaluate
$\cD$ perturbatively, then we obtain singular result of Refs. \cite{mak65} and \cite{gut98}.
The bound state removes the the singularity in the transmission probability obtained in perturbation theory, and leads to a thermal conductance that 
is finite and vanishes for $D\rightarrow 0$, but is nonanalytic in the normal-state transmission probability in the ``tunneling limit'' , 
i.e. $\kappa \sim D\ln D$.
The result for $\kappa(\vartheta)$ to leading order in $D$ also has non-analytic corrections to the phase modulation of the thermal conductance, in 
addition to the modulation $\propto\cos\vartheta$ (or equivalently $\sin^2(\vartheta/2)$),
\begin{equation}
\kappa(\vartheta)=\kappa_0-\kappa_1\sin^2{\vartheta\over 2}\ln(\sin^2{\vartheta\over 2})+\kappa_2 \sin^2{\vartheta\over 2}
\,,
\label{eq_conductance_small_D}
\end{equation}
where the coefficients are given in Ref. \cite{zha03}, and the relative importance of the $\ln$-term increases significantly for tunnel junctions ($D\ll 1$) at 
very low temperatures. 

In summary, phase-sensitive heat transport arises in Josephson point contacts from both the coherence factors for the Bogoliubov excitations that transport heat as well as the back-action of the Andreev bound-state spectrum on the transmission probability for these excitations. In the tunneling limit the shallow bound state acts as \emph{Andreev's Demon} by controlling the heat transport by tuning the resonance with the phase bias. This is shown in the right panel of Fig. \ref{fig_heat-conductance-Josephson-Point-Contact}, where the heat current is suppressed for $\vartheta =0$, but strongly enhanced by tuning the phase to $\vartheta=\pi$ to maximize resonant transmission of hot carriers near the gap edge.

\section{Outlook}

Experimental studies of charge transport in Josephson junctions, and the development of quantum devices based on phase-sensitive charge transport have a long history since Josephson's original prediction. By contrast  experimental studies, in particular the 
observation of phase-sensitive heat transport is relatively new. 
Quantum oscillations of the heat current in an SNS Andreev interferometer were reported in Ref. \cite{jia05}, and in a Josephson junction based interferomenter in 
Ref. \cite{gia12}. 
The experimental realization of quantum oscillations in heat transport has lead to substantial interest in the heat transport in mesoscopic superconducting devices based for phase-sensitive control of heat transport as a frontier research area \cite{gia06,for16}. 

In the broader context of unconventional and topological superconductors, charge, heat and spin transport provide wide-ranging opportunites for new physics, and potentially new device functionalities. Many of these frontier research areas are highlighted in this special volume. 

\dataccess This article has no additional data.
\competing I declare I have no competing interests. 
\funding{The research of the author has been supported by the National Science Foundation, and currently by Grants DMR-1508730 and PHY-1734332.}

\ack{I thank my collaborators, 
Matthias Eschrig, 
Micke Fogelstr\"om, 
Tomas L\"ofwander, 
Wave Ngampruetikorn,
Takeshi Mizushima,
Oleksii Shevtsov,
Anton Vorontsov,
Mehdi Zarea, 
Erhai Zhao, 
and the late Dierk Rainer, who was an inspiration for many theoretical developments on 
Andreev scattering and bound-state formation in unconventional superconductors.
}

\end{document}